\documentstyle[10pt,epsf,epsfig,hangcaption,xspace,amssymb,amsfonts,amsmath,amsthm,cite,
dp_delphititle,lineno]{dp_delphi}
\makeindex
\def\DpPaperGroup{EP}
\def\DpPaperRef{2002-090}
\def\DpDate{17 December 2002}
\def\DpAuthors{DELPHI Collaboration}
\def\DpSubmit{(Accepted by E. Phys. J. C )}
\def\DpTitle{Search for resonant $\tilde\nu$ production\\
at $\sqrt{s}=$ 183 to 208 GeV}
\def\DpComment{}
\def\DpEMail{}
\newcommand{\gev}{{\ifmmode \mbox{Ge\kern-0.2exV}
\else Ge\kern-0.2exV\nolinebreak\fi}}
\newcommand{\mev}{{\ifmmode \mbox{Me\kern-0.2exV}
\else Me\kern-0.2exV\nolinebreak\fi}}

\begin{document}
%%%%%%%%%%%%%%%%%%%%%%%%%% They are a problem with Coll.Sty ?
\makeatletter
%\input{dp_system:coll.sty}
% Collapse citation numbers to ranges.  Non-numeric and undefined labels
% are handled.  No sorting is done.  E.g., 1,3,2,3,4,5,foo,1,2,3,?,4,5
% gives 1,3,2-5,foo,1-3,?,4,5
\newcount\@tempcntc
\def\@citex[#1]#2{\if@filesw\immediate\write\@auxout{\string\citation{#2}}\fi
  \@tempcnta\z@\@tempcntb\m@ne\def\@citea{}\@cite{\@for\@citeb:=#2\do
    {\@ifundefined
       {b@\@citeb}{\@citeo\@tempcntb\m@ne\@citea\def\@citea{,}{\bf ?}\@warning
       {Citation `\@citeb' on page \thepage \space undefined}}%
    {\setbox\z@\hbox{\global\@tempcntc0\csname b@\@citeb\endcsname\relax}%
     \ifnum\@tempcntc=\z@ \@citeo\@tempcntb\m@ne
       \@citea\def\@citea{,}\hbox{\csname b@\@citeb\endcsname}%
     \else
      \advance\@tempcntb\@ne
      \ifnum\@tempcntb=\@tempcntc
      \else\advance\@tempcntb\m@ne\@citeo
      \@tempcnta\@tempcntc\@tempcntb\@tempcntc\fi\fi}}\@citeo}{#1}}
\def\@citeo{\ifnum\@tempcnta>\@tempcntb\else\@citea\def\@citea{,}%
  \ifnum\@tempcnta=\@tempcntb\the\@tempcnta\else
   {\advance\@tempcnta\@ne\ifnum\@tempcnta=\@tempcntb \else \def\@citea{--}\fi
    \advance\@tempcnta\m@ne\the\@tempcnta\@citea\the\@tempcntb}\fi\fi}
 
\makeatother
%%%%%%%%%%%%%%%%%%%%%%%%%% ??????????????????????????????????
% Generate the title page
\begin{titlepage}
\pagenumbering{roman}
\CERNpreprint{\DpPaperGroup}{\DpPaperRef} % Reference of the paper
\date{{\small\DpDate}} % Date of the paper
\title{\DpTitle} % Title of the paper
\address{\DpAuthors} % General name of the author(s)
\begin{shortabs} % Start the abstract
\noindent
\noindent

Searches for resonant $\tilde\nu$ production in
$e^+e^-$ collisions
under the assumption that
R-parity is not conserved and that the dominant R-parity violating coupling
is $\lambda_{121}$ or $\lambda_{131}$
used data recorded by DELPHI in 1997 to 2000
at centre-of-mass energies of 183 to 208~GeV.
No deviation from the Standard Model
was observed. Upper limits are given for
the $\lambda_{121}$ and $\lambda_{131}$ couplings as a function 
of the sneutrino
mass and total width. The limits are especially stringent
for sneutrino masses 
%around 183, 189, 196, 200 and 206~GeV/$c^2$, corresponding 
equal to the centre-of-mass energies with the highest
integrated luminosities recorded.
\end{shortabs}
\vfill
\begin{center}
\DpSubmit \ \\ % Horrible hack to allow to have DpSubmit empty
\DpComment \ \\
\DpEMail \ \\
\end{center}
\vfill
\clearpage
\headsep 10.0pt
\addtolength{\textheight}{10mm}
\addtolength{\footskip}{-5mm}
\begingroup
% Commands to process the author names
%
\newcommand{\DpName}[2]{\hbox{#1$^{\ref{#2}}$},\hfill}
\newcommand{\DpNameTwo}[3]{\hbox{#1$^{\ref{#2},\ref{#3}}$},\hfill}
\newcommand{\DpNameThree}[4]{\hbox{#1$^{\ref{#2},\ref{#3},\ref{#4}}$},\hfill}
\newskip\Bigfill \Bigfill = 0pt plus 1000fill
\newcommand{\DpNameLast}[2]{\hbox{#1$^{\ref{#2}}$}\hspace{\Bigfill}}
\small
%\footnotesize
\noindent
\DpName{J.Abdallah}{LPNHE}
\DpName{P.Abreu}{LIP}
\DpName{W.Adam}{VIENNA}
\DpName{P.Adzic}{DEMOKRITOS}
\DpName{T.Albrecht}{KARLSRUHE}
\DpName{T.Alderweireld}{AIM}
\DpName{R.Alemany-Fernandez}{CERN}
\DpName{T.Allmendinger}{KARLSRUHE}
\DpName{P.P.Allport}{LIVERPOOL}
\DpName{U.Amaldi}{MILANO2}
% ...

\normalsize
\endgroup
\titlefoot{Department of Physics and Astronomy, Iowa State
     University, Ames IA 50011-3160, USA
    \label{AMES}}
\titlefoot{Physics Department, Universiteit Antwerpen,
     Universiteitsplein 1, B-2610 Antwerpen, Belgium \\
     \indent~~and IIHE, ULB-VUB,
     Pleinlaan 2, B-1050 Brussels, Belgium \\
     \indent~~and Facult\'e des Sciences,
     Univ. de l'Etat Mons, Av. Maistriau 19, B-7000 Mons, Belgium
    \label{AIM}}
\titlefoot{Physics Laboratory, University of Athens, Solonos Str.
     104, GR-10680 Athens, Greece
    \label{ATHENS}}
\titlefoot{Department of Physics, University of Bergen,
     All\'egaten 55, NO-5007 Bergen, Norway
    \label{BERGEN}}
\titlefoot{Dipartimento di Fisica, Universit\`a di Bologna and INFN,
     Via Irnerio 46, IT-40126 Bologna, Italy
    \label{BOLOGNA}}
\titlefoot{Centro Brasileiro de Pesquisas F\'{\i}sicas, rua Xavier Sigaud 150,
     BR-22290 Rio de Janeiro, Brazil \\
     \indent~~and Depto. de F\'{\i}sica, Pont. Univ. Cat\'olica,
     C.P. 38071 BR-22453 Rio de Janeiro, Brazil \\
     \indent~~and Inst. de F\'{\i}sica, Univ. Estadual do Rio de Janeiro,
     rua S\~{a}o Francisco Xavier 524, Rio de Janeiro, Brazil
    \label{BRASIL}}
\titlefoot{Coll\`ege de France, Lab. de Physique Corpusculaire, IN2P3-CNRS,
     FR-75231 Paris Cedex 05, France
    \label{CDF}}
\titlefoot{CERN, CH-1211 Geneva 23, Switzerland
    \label{CERN}}
\titlefoot{Institut de Recherches Subatomiques, IN2P3 - CNRS/ULP - BP20,
     FR-67037 Strasbourg Cedex, France
    \label{CRN}}
\titlefoot{Now at DESY-Zeuthen, Platanenallee 6, D-15735 Zeuthen, Germany
    \label{DESY}}
\titlefoot{Institute of Nuclear Physics, N.C.S.R. Demokritos,
     P.O. Box 60228, GR-15310 Athens, Greece
    \label{DEMOKRITOS}}
\titlefoot{FZU, Inst. of Phys. of the C.A.S. High Energy Physics Division,
     Na Slovance 2, CZ-180 40, Praha 8, Czech Republic
    \label{FZU}}
\titlefoot{Dipartimento di Fisica, Universit\`a di Genova and INFN,
     Via Dodecaneso 33, IT-16146 Genova, Italy
    \label{GENOVA}}
\titlefoot{Institut des Sciences Nucl\'eaires, IN2P3-CNRS, Universit\'e
     de Grenoble 1, FR-38026 Grenoble Cedex, France
    \label{GRENOBLE}}
\titlefoot{Helsinki Institute of Physics, HIP,
     P.O. Box 9, FI-00014 Helsinki, Finland
    \label{HELSINKI}}
\titlefoot{Joint Institute for Nuclear Research, Dubna, Head Post
     Office, P.O. Box 79, RU-101 000 Moscow, Russian Federation
    \label{JINR}}
\titlefoot{Institut f\"ur Experimentelle Kernphysik,
     Universit\"at Karlsruhe, Postfach 6980, DE-76128 Karlsruhe,
     Germany
    \label{KARLSRUHE}}
\titlefoot{Institute of Nuclear Physics,Ul. Kawiory 26a,
     PL-30055 Krakow, Poland
    \label{KRAKOW1}}
\titlefoot{Faculty of Physics and Nuclear Techniques, University of Mining
     and Metallurgy, PL-30055 Krakow, Poland
    \label{KRAKOW2}}
\titlefoot{Universit\'e de Paris-Sud, Lab. de l'Acc\'el\'erateur
     Lin\'eaire, IN2P3-CNRS, B\^{a}t. 200, FR-91405 Orsay Cedex, France
    \label{LAL}}
\titlefoot{School of Physics and Chemistry, University of Lancaster,
     Lancaster LA1 4YB, UK
    \label{LANCASTER}}
\titlefoot{LIP, IST, FCUL - Av. Elias Garcia, 14-$1^{o}$,
     PT-1000 Lisboa Codex, Portugal
    \label{LIP}}
\titlefoot{Department of Physics, University of Liverpool, P.O.
     Box 147, Liverpool L69 3BX, UK
    \label{LIVERPOOL}}
\titlefoot{LPNHE, IN2P3-CNRS, Univ.~Paris VI et VII, Tour 33 (RdC),
     4 place Jussieu, FR-75252 Paris Cedex 05, France
    \label{LPNHE}}
\titlefoot{Department of Physics, University of Lund,
     S\"olvegatan 14, SE-223 63 Lund, Sweden
    \label{LUND}}
\titlefoot{Universit\'e Claude Bernard de Lyon, IPNL, IN2P3-CNRS,
     FR-69622 Villeurbanne Cedex, France
    \label{LYON}}
\titlefoot{Dipartimento di Fisica, Universit\`a di Milano and INFN-MILANO,
     Via Celoria 16, IT-20133 Milan, Italy
    \label{MILANO}}
\titlefoot{Dipartimento di Fisica, Univ. di Milano-Bicocca and
     INFN-MILANO, Piazza della Scienza 2, IT-20126 Milan, Italy
    \label{MILANO2}}
\titlefoot{IPNP of MFF, Charles Univ., Areal MFF,
     V Holesovickach 2, CZ-180 00, Praha 8, Czech Republic
    \label{NC}}
\titlefoot{NIKHEF, Postbus 41882, NL-1009 DB
     Amsterdam, The Netherlands
    \label{NIKHEF}}
\titlefoot{National Technical University, Physics Department,
     Zografou Campus, GR-15773 Athens, Greece
    \label{NTU-ATHENS}}
\titlefoot{Physics Department, University of Oslo, Blindern,
     NO-0316 Oslo, Norway
    \label{OSLO}}
\titlefoot{Dpto. Fisica, Univ. Oviedo, Avda. Calvo Sotelo
     s/n, ES-33007 Oviedo, Spain
    \label{OVIEDO}}
\titlefoot{Department of Physics, University of Oxford,
     Keble Road, Oxford OX1 3RH, UK
    \label{OXFORD}}
\titlefoot{Dipartimento di Fisica, Universit\`a di Padova and
     INFN, Via Marzolo 8, IT-35131 Padua, Italy
    \label{PADOVA}}
\titlefoot{Rutherford Appleton Laboratory, Chilton, Didcot
     OX11 OQX, UK
    \label{RAL}}
\titlefoot{Dipartimento di Fisica, Universit\`a di Roma II and
     INFN, Tor Vergata, IT-00173 Rome, Italy
    \label{ROMA2}}
\titlefoot{Dipartimento di Fisica, Universit\`a di Roma III and
     INFN, Via della Vasca Navale 84, IT-00146 Rome, Italy
    \label{ROMA3}}
\titlefoot{DAPNIA/Service de Physique des Particules,
     CEA-Saclay, FR-91191 Gif-sur-Yvette Cedex, France
    \label{SACLAY}}
\titlefoot{Instituto de Fisica de Cantabria (CSIC-UC), Avda.
     los Castros s/n, ES-39006 Santander, Spain
    \label{SANTANDER}}
\titlefoot{Inst. for High Energy Physics, Serpukov
     P.O. Box 35, Protvino, (Moscow Region), Russian Federation
    \label{SERPUKHOV}}
\titlefoot{J. Stefan Institute, Jamova 39, SI-1000 Ljubljana, Slovenia
     and Laboratory for Astroparticle Physics,\\
     \indent~~Nova Gorica Polytechnic, Kostanjeviska 16a, SI-5000 Nova Gorica, Slovenia, \\
     \indent~~and Department of Physics, University of Ljubljana,
     SI-1000 Ljubljana, Slovenia
    \label{SLOVENIJA}}
\titlefoot{Fysikum, Stockholm University,
     Box 6730, SE-113 85 Stockholm, Sweden
    \label{STOCKHOLM}}
\titlefoot{Dipartimento di Fisica Sperimentale, Universit\`a di
     Torino and INFN, Via P. Giuria 1, IT-10125 Turin, Italy
    \label{TORINO}}
\titlefoot{INFN,Sezione di Torino, and Dipartimento di Fisica Teorica,
     Universit\`a di Torino, Via P. Giuria 1,\\
     \indent~~IT-10125 Turin, Italy
    \label{TORINOTH}}
\titlefoot{Dipartimento di Fisica, Universit\`a di Trieste and
     INFN, Via A. Valerio 2, IT-34127 Trieste, Italy \\
     \indent~~and Istituto di Fisica, Universit\`a di Udine,
     IT-33100 Udine, Italy
    \label{TU}}
\titlefoot{Univ. Federal do Rio de Janeiro, C.P. 68528
     Cidade Univ., Ilha do Fund\~ao
     BR-21945-970 Rio de Janeiro, Brazil
    \label{UFRJ}}
\titlefoot{Department of Radiation Sciences, University of
     Uppsala, P.O. Box 535, SE-751 21 Uppsala, Sweden
    \label{UPPSALA}}
\titlefoot{IFIC, Valencia-CSIC, and D.F.A.M.N., U. de Valencia,
     Avda. Dr. Moliner 50, ES-46100 Burjassot (Valencia), Spain
    \label{VALENCIA}}
\titlefoot{Institut f\"ur Hochenergiephysik, \"Osterr. Akad.
     d. Wissensch., Nikolsdorfergasse 18, AT-1050 Vienna, Austria
    \label{VIENNA}}
\titlefoot{Inst. Nuclear Studies and University of Warsaw, Ul.
     Hoza 69, PL-00681 Warsaw, Poland
    \label{WARSZAWA}}
\titlefoot{Fachbereich Physik, University of Wuppertal, Postfach
     100 127, DE-42097 Wuppertal, Germany \\
\noindent
{$^\dagger$~deceased}
    \label{WUPPERTAL}}

\addtolength{\textheight}{-10mm}
\addtolength{\footskip}{5mm}
\clearpage
\headsep 30.0pt
\end{titlepage}
%%%%%%%%%%%%%%%%%%%%%%%%%
%
% Change for the document body
%%\pagestyle{heading} % for page numbering
\pagenumbering{arabic} % page numbering in number
\setcounter{footnote}{0} %
\large
%\linenumbers %%%CD
\section{Introduction}

In the Minimal Supersymmetric extension of the Standard 
Model~(MSSM)~\cite{mssm}, a discrete symmetry called 
R-parity~\cite{fayet}
predicts baryon number ($B$) and lepton number ($L$) conservation
which is an accidental feature of the 
SU(3)$\times$SU(2)$\times$U(1) Standard Model (SM).
The related quantum number $R_p=(-1)^{3B+L+2S}$,
where $S$ is the spin of the particle, is multiplicatively conserved.
However, from a theoretical point of view,
R-parity conservation is not needed. Allowing its violation
leads to a more general superpotential $W$ which can 
include the following renormalizable gauge invariant
additional terms:
$$ W_{\Delta L\neq 0}=
\lambda _{ijk}L_iL_j\bar E_k
+\lambda ^\prime_{ijk}L_iQ_j\bar D_k
+\epsilon _iH_uL_i$$
$$ W_{\Delta B\neq 0}=
\lambda ^{\prime\prime}_{ijk}\bar U_i\bar D_j\bar D_k.$$
Here $L$ ($Q$) are the lepton (quark) doublet superfields,
$\bar E$ ($\bar U$, $\bar D$) are the lepton (up and down quark)
singlet superfields, $H_u$ is the Higgs superdoublet coupling
to up-type quarks and leptons,
$i$, $j$, $k$ are generation indices;
$\epsilon _i$ are parameters with dimensions while
$\lambda _{ijk}$, $\lambda ^\prime_{ijk}$,
$\lambda ^{\prime\prime}_{ijk}$ are dimensionless Yukawa-like couplings.
The $\lambda _{ijk}$ ($\lambda ^{\prime\prime}_{ijk}$) couplings
are anti-symmetric in the first (last) two indices because
of gauge invariance.

Nevertheless, it is necessary to require that $\Delta L\neq 0$ and
$\Delta B\neq 0$ terms are not both present to avoid
a fast proton decay. In this paper, it will be assumed
that only one $\lambda _{ijk}$ coupling is non-vanishing.

\begin{figure}
\begin{center}
\epsfig{file=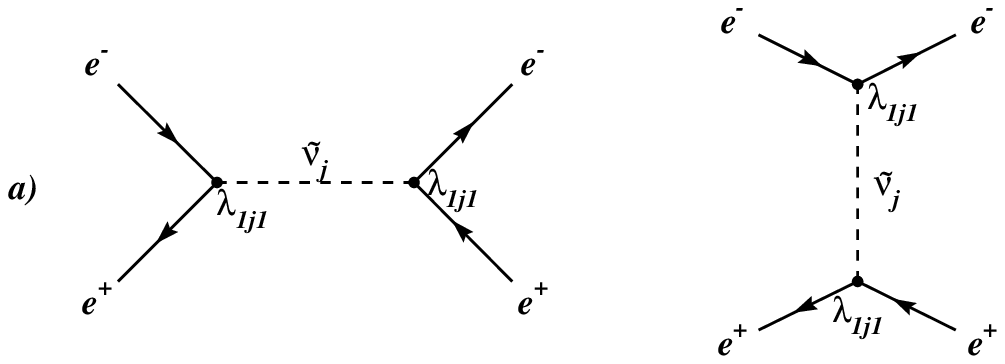,width=14cm}\\
\epsfig{file=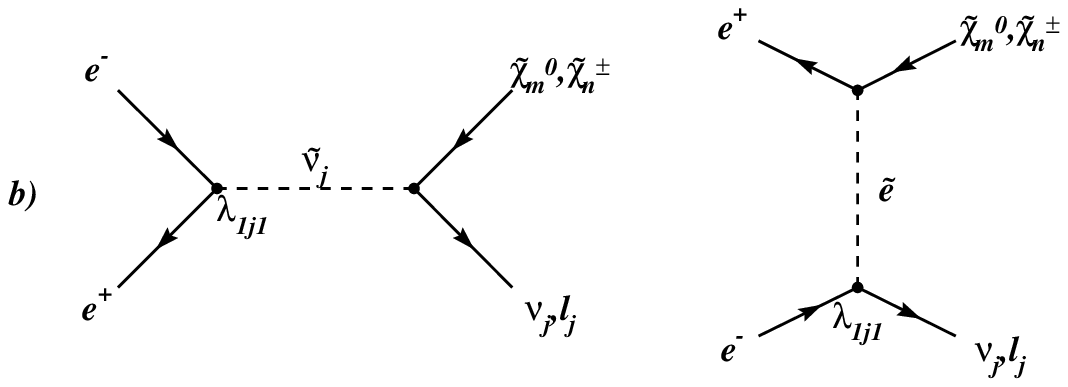,width=14cm}
\end{center}
\caption{Lowest order Feynman diagrams for single sneutrino production
and decay: a) direct decay, b) indirect decay
($\nu _j=\nu _\mu, \nu _\tau$;
$l _j=\mu, \tau$; $m=1$,4; $n=1$,2).}
\label{fig:diag}
\end{figure}

R-parity violation has two major consequences. It allows
the decay of the lightest supersymmetric particle (LSP),
thus discarding it as a candidate to cold dark matter.
It also allows the supersymmetric particles to be singly produced,
via $\lambda _{ijk}$ couplings in the case of $e^+e^-$ collisions.
It is this possibility that is explored in this paper.
If $\lambda _{121}$ or $\lambda _{131}$ is non-vanishing, a muon
sneutrino or a tau sneutrino (an electron sneutrino cannot be produced
because $\lambda _{111}=0$), with spin 0,
can be produced in the $s$-channel~\cite{hall} (see Figure~\ref{fig:diag}). 
The simplest expression for the cross-section is~\cite{barger}:
 $$ \sigma (e^+ e^- \to \tilde\nu  \rightarrow  X ) (s) = 
 \displaystyle{\frac{4 \pi s}{M_{\tilde\nu}^2}}
 \displaystyle{\frac{\Gamma (ee) \Gamma (X)}
 { (s- M_{\tilde\nu}^2)^2 + M_{\tilde\nu}^2 \Gamma_{\tilde\nu } ^2 }}  $$
 where $ \Gamma (ee) = \Gamma (\tilde\nu _j \rightarrow e^+ e^-) 
= \displaystyle{\frac{\lambda_{1j1}^2}{ 16 \pi} M_{\tilde\nu}}$,
$j=2$, 3
and $ \Gamma (X)$  denotes the partial width for $\tilde\nu _j$ 
decay to a final state $X$, with $X = e^+ e^-$ (sneutrino direct decay), 
$\tilde\chi ^0 \nu$ or $\tilde\chi ^\pm l^{\mp}$ (sneutrino indirect
decays). In the indirect decay mode, 
$\Gamma (X)$ is independent of $\lambda _{1j1}$.
The additional $t$-channel contributions and the interference terms,
not shown in this formula for reasons of
compactness, were included in the signal
simulation~\cite{susygen} and final estimation of expected events.
This cross-section is expected to be
very high for $M_{\tilde\nu}\simeq \sqrt{s}$ 
(of the order of 50~pb for instance for $\lambda _{1j1}=10^{-2}$)
and to remain large for masses below the centre-of-mass energy,
due to initial state radiation effects, again not displayed by
this formula but taken into account in the analysis.

  Given the present (indirect) upper limits on $\lambda_{121}$ and 
$\lambda_{131}$ ($\lambda_{121} < 0.04
\times \frac{M_{\tilde e_R}}{100 \mbox{\tiny \ GeV/}c^2}$,  
$\lambda_{131} < 0.05\times \frac{M_{\tilde e_R}}{100 \mbox{ \tiny GeV/}c^2}$ 
at 68\% confidence level (CL)~\cite{indir}),
the $ e^+ e^-$ decay channel, having a cross-section proportional to 
${\lambda}^4$, is suppressed compared 
to the other two ($\sigma\propto {\lambda}^2$), unless all
neutralinos and charginos are heavier than the sneutrino.
The direct decay mode has already been investigated by
the LEP collaborations~\cite{direct} by looking for deviations
from the Standard Model in the cross-sections and asymmetries 
of $e^+e^-\to l^+l^-$. The results were 
%obtained assuming a 1~GeV/$c^2 total width ($\Gamma _{\tilde\nu}$)
%for the sneutrino and 
presented as an upper limit
on $\lambda _{1j1}$ as a function of the sneutrino mass $M_{\tilde\nu}$.
The indirect decay modes are analysed explicitly here,
taking into account any mass and width of the sneutrino
as a function of the MSSM parameters. The scheme chosen here
is a constrained MSSM in which the SUSY breaking occurs
via gravitational interactions
%is a minimal super-gravity 
(mSUGRA \cite{mssm}). 
The relevant parameters are then: 
$M_2$,
the SU(2) gaugino mass at the electroweak scale;
$m_0$,
the scalars common mass at the Grand Unified Theories (GUT) scale;
$\mu$,
the mixing mass term of the Higgs doublets at the electroweak scale;
and $\tan{\beta}$, the ratio of the vacuum expectation values
of the two Higgs doublets. 
The unified trilinear coupling $A_0$ is assumed to be zero.
It is also assumed that the running of the
$\lambda$ couplings from the GUT
scale to the electroweak scale
does not have a significant effect on the running of the
gaugino and sfermion masses.

%The phenomenology of the MSSM with R-parity violation
%via $\lambda$ couplings
%has been described in detail in a previous paper~\cite{corinne}
%in which pair production was studied.

In this model, the LSP is generally 
the lightest neutralino $\tilde\chi _1^0$;
with a $\lambda _{121}$ ($\lambda _{131}$) coupling,
it decays to $e\mu\nu _e$ or $ee\nu _\mu$ 
($e\tau\nu _e$ or $ee\nu _\tau$).
All other sparticles, like the sneutrino, can have both
direct and indirect decays. In particular, the lightest
chargino  $\tilde\chi _1^+$ can decay either to $ee\mu$, $e\nu _e\nu _\mu$
($\lambda _{121}$) or to the $R_p$ conserving channel 
$\tilde\chi ^0_1 W^\star$. The latter is generally clearly
dominant over the former, unless the $\lambda$ coupling is very large
or the mass difference between the chargino and the neutralino
is very small. The decays of the heavier neutralinos and charginos
are similar, with the additional possibility of 
longer cascade decays always leading to the LSP decay.

A complete review of the possible decays of the neutralinos
and charginos showed that the final states of all indirect decays
of the sneutrino could be classified into three topologies (or channels):
\begin{itemize}
\item[1)] events with two leptons and missing energy;
\item[2)] events with four or six leptons (with or without missing energy);
\item[3)] events with at least two isolated leptons and at least two hadronic jets.
\end{itemize}
The first topology mostly comes from 
$\tilde\nu\to\tilde\chi _1^0\nu$, $\tilde\chi _1^0\to e^\pm l^\mp\nu$,
leading to a $2l+\nu$ final state.
The second one is dominated by
$\tilde\nu\to\tilde\chi _1^+l^-$, $\tilde\chi _1^+\to \tilde\chi _1^0 l^+\nu$,
$\tilde\chi _1^0\to e^\pm l^\mp\nu$, leading to $4l+2\nu$.
The third topology, being relatively general, often has a high branching ratio.
It can arise for example from 
$\tilde\nu\to\tilde\chi _1^+l^-$, 
$\tilde\chi _1^+\to \tilde\chi _1^0 q\bar q^\prime$,
$\tilde\chi _1^0\to e^\pm l^\mp\nu$, leading to $3l+2$jets$+\nu$,
or from  
$\tilde\nu\to\tilde\chi _2^0\nu$, 
$\tilde\chi _2^0\to \tilde\chi _1^0 q\bar q$,
$\tilde\chi _1^0\to e^\pm l^\mp\nu$, leading to $2l+2$jets$+2\nu$.\\

The paper is organised as follows. Section 2 lists the data
samples that were used in the present search for resonant
sneutrino production. In section 3, the selection criteria
are described, as well as the results of the selection.
Finally, in section 4 
the limits on the $\lambda _{1j1}$ couplings 
are derived by comparing the Standard Model expectations
and the experimental results.

\section{Data samples}

A detailed description of the DELPHI detector can be found in~\cite{detec}.
The present analysis was mainly based on the capability of reconstructing
charged particle tracks using the tracking devices, particularly
the Time Projection Chamber (TPC) but also the silicon Vertex Detector,
the drift chambers called Inner and Outer Detectors, and the forward detectors.
The complete system was inside a solenoidal magnetic field
of 1.2~T, parallel to the beam axis.
The analysis also used the lepton identification
capabilities of the electromagnetic calorimeters (the barrel High
density Projection Chamber HPC and the Forward Electro-Magnetic
Calorimeter FEMC) for the electrons and of the muon chambers
for the muons. The hadron calorimeter was used
to detect neutral hadrons.

The data from 1997 to 2000 LEP runs were taken at centre-of-mass
energies between 183~GeV and 208~GeV. The registered 
integrated luminosities %$\int\!{\cal L}$
after requiring the TPC and all the calorimeters 
(HPC, FEMC and hadron calorimeter)
to be operational are given in Table~\ref{tab:lumi}.
This quality requirement rejected 2\% of the 
luminosity recorded in 1997 to 1999 and 29\% 
of 2000 integrated luminosity, because one of the
twelve sectors of the TPC was off before the end of the data taking.
Runs with partly inefficient muon chambers were kept,
and the simulation was adjusted to reproduce
the effective efficiency.\\

\begin{table}
\begin{center}
\begin{tabular}{|c|c|}
\hline
$<\! \sqrt{s}\! >$(GeV) & ${\int\!\cal L}$ (pb$^{-1}$) \\
\hline
182.7 & \hfill 52.2~~~~  \\
188.6 & \hfill 153.8~~~~  \\
191.6 & \hfill 25.1~~~~  \\
195.5 & \hfill 75.9~~~~ \\
199.5 & \hfill 82.8~~~~ \\
201.6 & \hfill 43.2~~~~ \\
203.7 & \hfill 6.3~~~~ \\
205.0 & \hfill 67.2~~~~ \\
206.5 & \hfill 78.2~~~~ \\
208.0 & \hfill 7.3~~~~ \\
\hline
All & \hfill 592.0~~~~ \\
\hline
\end{tabular}
\caption{Average centre-of-mass energies and 
integrated luminosities.}
%${\int\!\cal L}$
\label{tab:lumi}
\end{center}
\end{table}

To evaluate the background contamination, different contributions
coming from the Standard Model were considered. The Standard Model events
were produced by the following generators: {\tt BDKRC}~\cite{bdk} for
the $e^+e^-\to e^+e^-l^+l^-$ four-fermion events of type $\gamma\gamma$,
and {\tt WPHACT}\cite{wph} for the other four-fermion events;
{\tt KK2F}~\cite{kk2f} for
the two-fermion events of type $e^+e^-\to f\bar f (\gamma )$,
with $f\neq e$, $\tau$, {\tt BHWIDE}~\cite{bhw} for the Bhabha events
($f=e$) and {\tt KORALZ}~\cite{kora} for $f=\tau$ events;
{\tt PYTHIA}~\cite{pyt} and {\tt WPHACT} for the $\gamma\gamma\to$ hadrons
events.

Signal events were generated with the {\tt SUSYGEN 2.20} 
generator~\cite{susygen}.
Samples of 3500 to 6000 events were generated
for nine MSSM parameter sets (see Table~\ref{tab:param}).\\

Simulated events were produced from the generated samples
with the standard DELPHI simulation program
{\tt DELSIM}~\cite{detec} 
and passed through the same reconstruction chain as the data.
These events were used to design the event selection. 

A faster simulation programme ({\tt SGV})~\cite{sgv} 
was applied to the same generated samples 
in order to be validated for further use in the limits extraction.
% voir validation de sgv p. 15 et 18 de mon cahier    

\begin{table}[bt]
\begin{center}
\begin{tabular}{|c|rrr|lcr|ccc|}
\hline
Parameter set & $m_0$ & $\mu$~ & $M_2$ & $\Gamma _{\tilde\nu}$ &
m($\tilde\chi _1^0$) & m($\tilde\chi _1^+$) &
Br$_1$ & Br$_2$ & Br$_3$\\
\hline
1 & 150 & 175 & 125 & 1 & 37 & 68~ & 0.39 & 0.23 & 0.38\\
2 & 190 & 275 & 155 & 1 & 65 & 116~ & 0.42 & 0.24 & 0.34\\
3 & 200 & 195 & 175 & 1 & 64 & 106~ & 0.41 & 0.20 & 0.38\\
4 & 207 & 125 & 385 & 0.2 & 87 & 104~ & 0.57 & 0.15 & 0.28\\
5 & 207 & -75 & 155 & 0.5 & 71 & 96~ & 0.19 & 0.24 & 0.57\\
6 & 207 & -125 & 115 & 1 & 63 & 127~ & 0.10 & 0.17 & 0.72\\
7 & 207 & 305 & 135 & 1.5 & 57 & 105~ & 0.37 & 0.24 & 0.39\\
8 & 207 & -285 & 65 & 2 & 36 & 81~ & 0.17 & 0.37 & 0.46\\
9 & 220 & 285 & 185 & 1 & 80 & 142~ & 0.52 & 0.18 & 0.31\\
\hline
\end{tabular}
\end{center}
\caption{Values of the SUSY parameters (in GeV/$c^2$)
used in the signal simulation, and resulting width, masses and branching ratios
(Br$_i$ is the branching ratio of channel number~$i$).}
%$\tan{\beta}=1.5$ in all cases.}
%in Figure~\ref{fig:sgv}.}
\label{tab:param}
\end{table}

\section{Event selection}

In order to select the three categories of final states defined above, 
a preselection was first applied.
In the following, charged particles reconstructed from their
trajectories in the tracking chambers were accepted only if their
momenta were greater than 100~MeV/$c$
and less than 1.5 times the beam energy. 
They also had to have a relative
momentum error less than 100\% and impact parameters
of at most 4~cm both in the plane transverse to the beam
and along the beam direction.
On the other hand, a neutral particle is assumed to be detected
as a cluster of energy deposits 
%(of at least 300~MeV in the HPC or 400~MeV in the FEMC) 
not associated to a track.
In each event, the following were required:

\begin{itemize}
\item at least two charged particles;
\item the sum of all charged particle energies %$E_{ch}$ 
      greater than $0.1\times \sqrt{s}$
      (and in any case, total energy of all particles
      $E_{tot}$ greater than 20~GeV);
\item the total momentum transverse to the beam
      %\footnote{$p_T^2=\Sigma (p_x^2+p_y^2)$}        $p_T$ 
      greater than 5~GeV/$c$;
\item the absolute value of total electric charge %$|Q_{ev}|$ 
      at most 1 if there are less than 7 charged particles;
\item at least one charged particle in the barrel (polar angle
      between 40$^\circ$ and 140$^\circ$);
\item the absolute value of the cosine of the polar angle of the missing
      momentum vector %$|\cos\theta_{miss}|$ 
      below 0.95 (or 0.9 in case of exactly two charged particles);
\item at least one isolated (i.e. with no other track
      in a $5^\circ$ half-cone centred on its direction) identified lepton
      (electron or muon)
      with momentum above 5~GeV/$c$ and with a maximum angle
      of 170$^\circ$ with respect to the nearest track.
\end{itemize}

The lepton identification used standard DELPHI tools~\cite{detec}.
The electron identification algorithm relied on two types of informations:
the energy deposited in the electromagnetic calorimeters and 
the dE/dx measurement in the TPC. 
The muon identification algorithm was based on the association
of signals in the muon chambers with extrapolated tracks; the most
efficient set of criteria, as described in ~\cite{detec},
was chosen for this analysis.

%The efficiency of these requirements is on average 
%at $\sqrt{s}\simeq 206$~GeV of 72\%, 90\% and 87\%
%respectively on the two-lepton, four or six lepton and semi-leptonic
%$\lambda_{121}$ signals, and of 62\%, 81\%, 75\% 
%on the $\lambda_{131}$ signals.\\ 

The above criteria define the preselection.
Then four more series of requirements were designed, in order to select
the different kinds of topologies.
For the two acoplanar\footnote{The acoplanarity is $180^\circ$ minus
      the angle between the transverse momenta of the two charged particles,
      or of the two reconstructed jets 
      (forcing the number of jets to be two 
      with the LUCLUS algorithm~\cite{luclus}) 
      if the event contains more particles.}
lepton and four or six lepton topologies, the
criteria were the same for both $\lambda_{121}$ and $\lambda_{131}$ 
couplings; for the semi-leptonic topology, two slightly different
selections were applied.

\begin{itemize}

\item For the two acoplanar lepton final states, it was required that:
\begin{itemize}
\item there be exactly two charged particles;
\item not both identified as muons;
\item the acoplanarity be above $40^\circ$;
\item the acollinearity\footnote{The acollinearity is 
      $180^\circ$ minus the angle between the two charged particles
      (or the two jets, see the definition of acoplanarity).}
      be above $50^\circ$;
\item the invariant mass of the two leptons\footnote{When
         the second lepton was not identified, it was assumed to be
         an electron.} %, $M_{ll}$, 
         be lower than $0.25\times \sqrt{s}$;
\item the angle %$\alpha _{ll}$ 
      between the two leptons be lower than 100$^\circ$.
\end{itemize}

\item For the four or six lepton final states, the requirements were:
\begin{itemize}
\item four or six charged particles;
\item at least two identified leptons (electrons or muons);
\item the resolution parameter of the Durham algorithm~\cite{durham}
      for which the event changes from four to three jets 
      $y_{34}$ greater than $10^{-4}$.
\end{itemize}

\item For the semi-leptonic final states, $\lambda_{121}$ coupling:
\begin{itemize}
\item at least 7 charged particles and at most 25;
\item at least two identified leptons 
      (electrons or muons);
\item the transverse momentum of the second most energetic lepton %$p_T(l_2)$ 
      had to be above $0.05\times \sqrt{s}$;
\item $y_{34}$ had to be greater than $10^{-3}$;
\item when the number of jets was forced to four, 
      at least two of them were required to be thin, 
      that is
      to have a total (track + neutral) multiplicity not exceeding~4.
\end{itemize}

\item For the semi-leptonic final states, $\lambda_{131}$ coupling:
\begin{itemize}
\item at least 7 charged particles and at most 25;
\item at least two identified leptons (electrons or muons),
      including at least one identified electron;
\item $y_{34}$ had to be greater than $10^{-3}$;
\item when the number of jets was forced to four, 
      at least two of them were required to be thin;
      %the number of jets
      %with a maximum total multiplicity of 4 was required to be
      %at least 2,
\item the missing energy ($\sqrt{s}-E_{tot}$) had to be
      greater than $0.25\times \sqrt{s}$.
\end{itemize}

\end{itemize}

\begin{figure}
\begin{center}
\epsfig{file=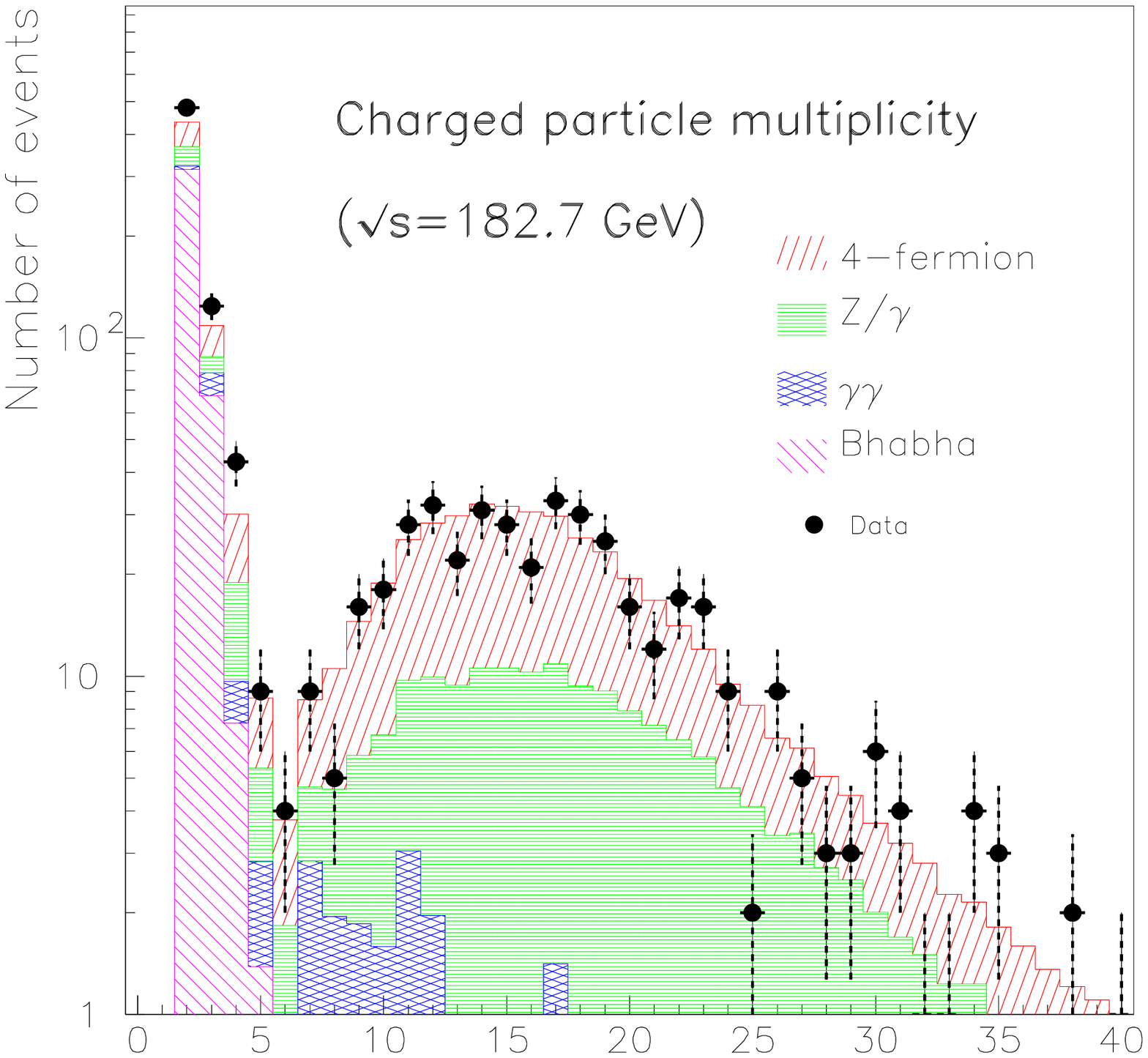,width=7.2cm}~\epsfig{file=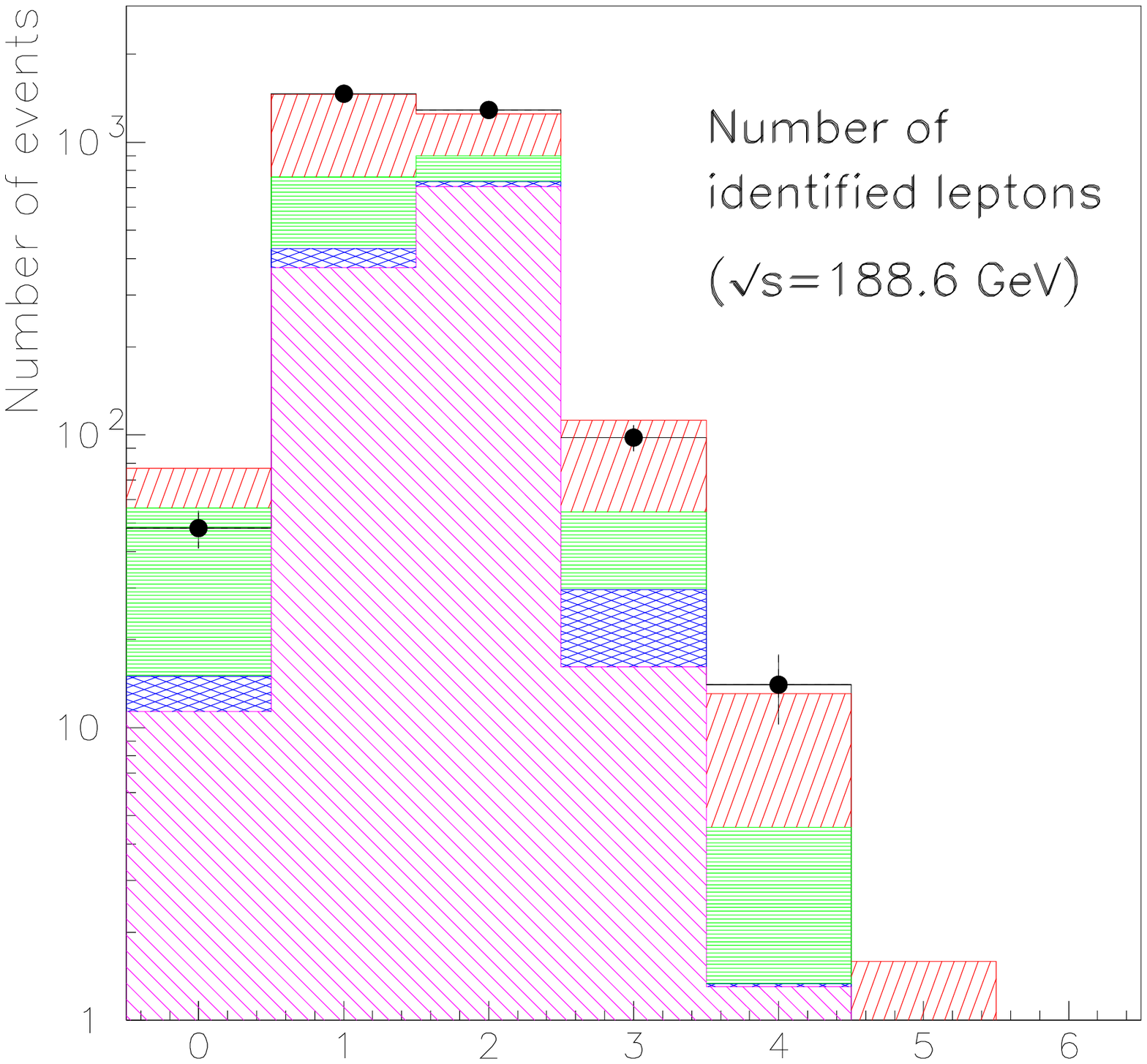,width=7.2cm}

\epsfig{file=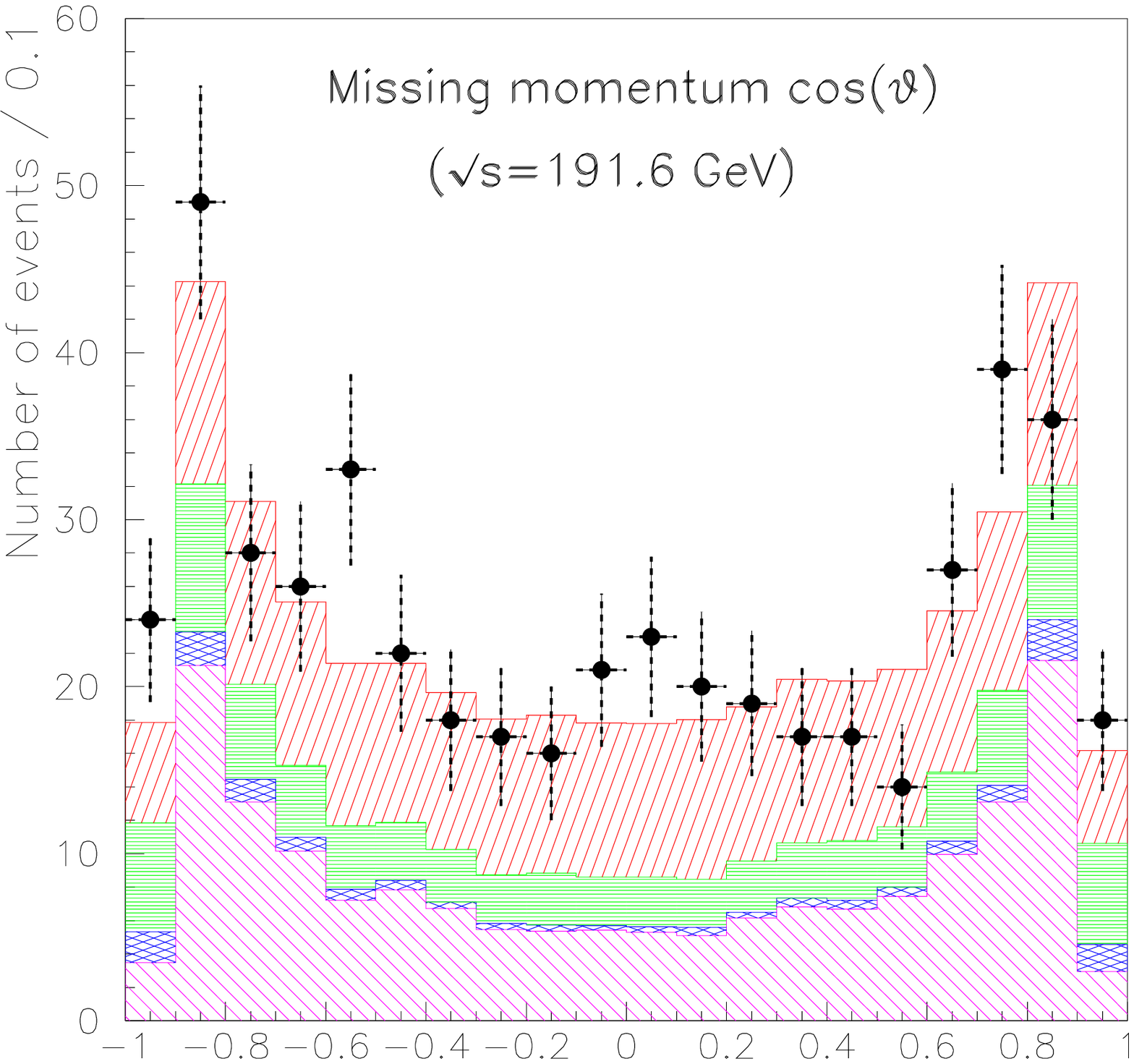,width=7.2cm}~\epsfig{file=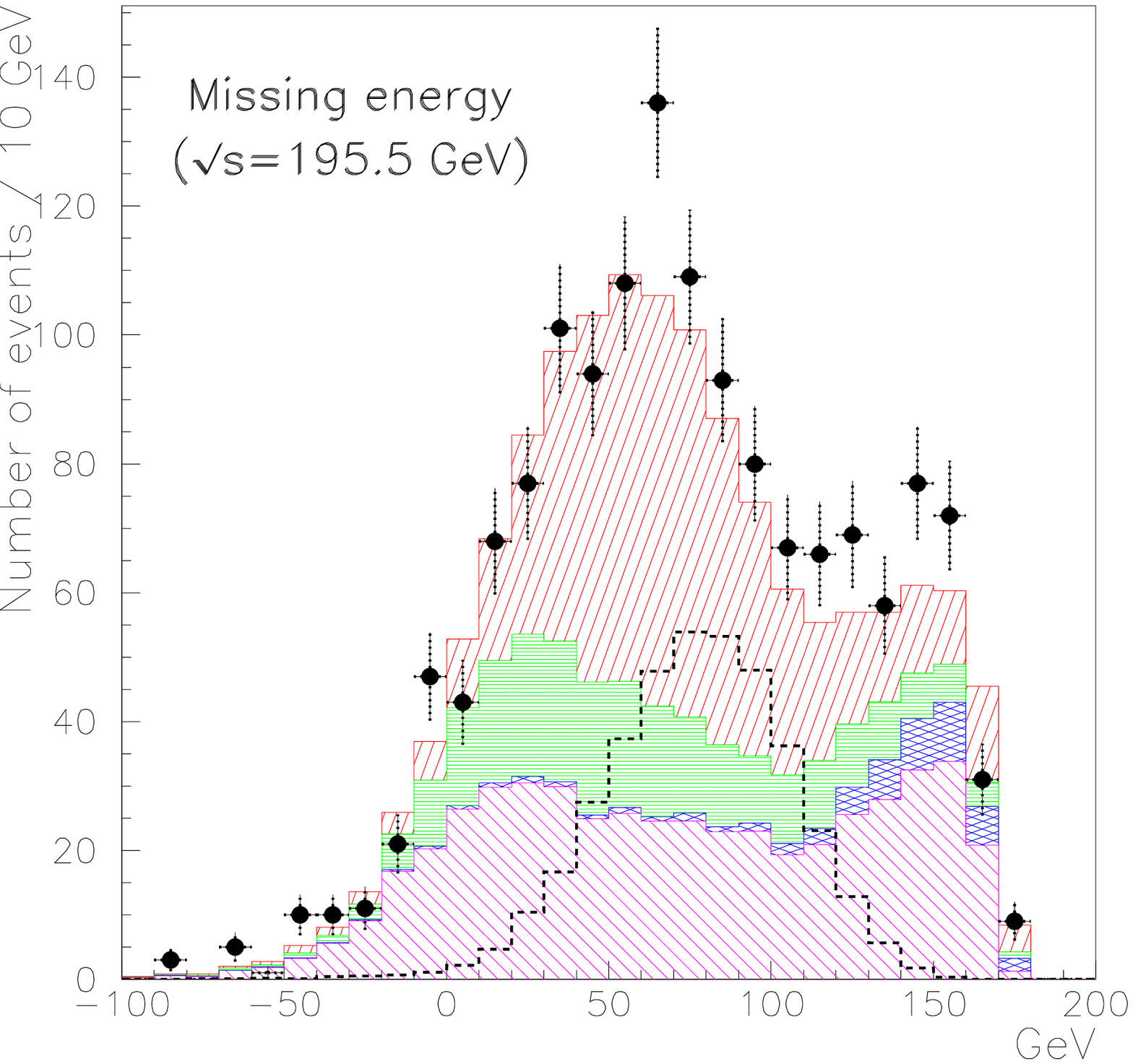,width=7.2cm}

\epsfig{file=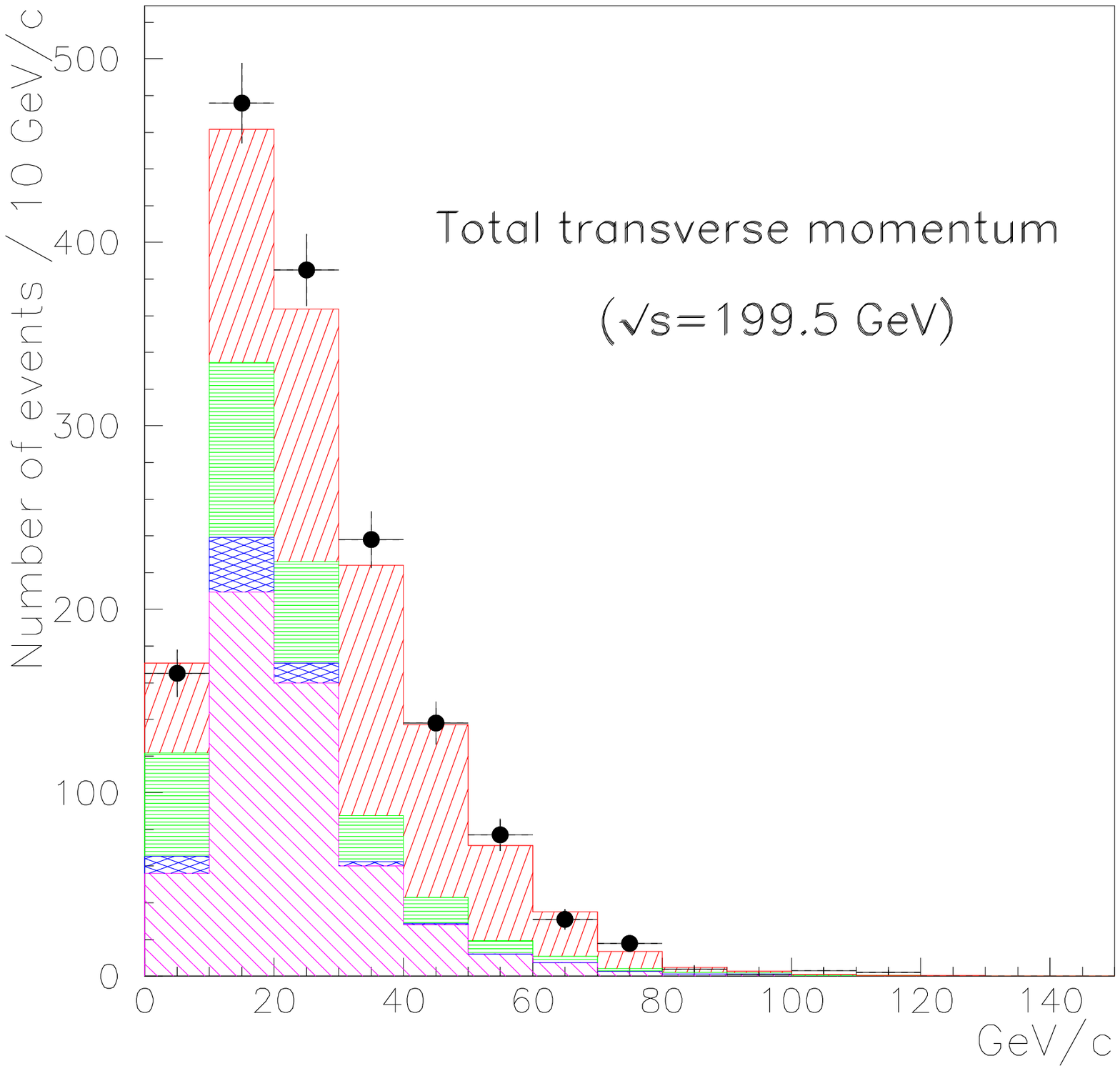,width=7.2cm}~\epsfig{file=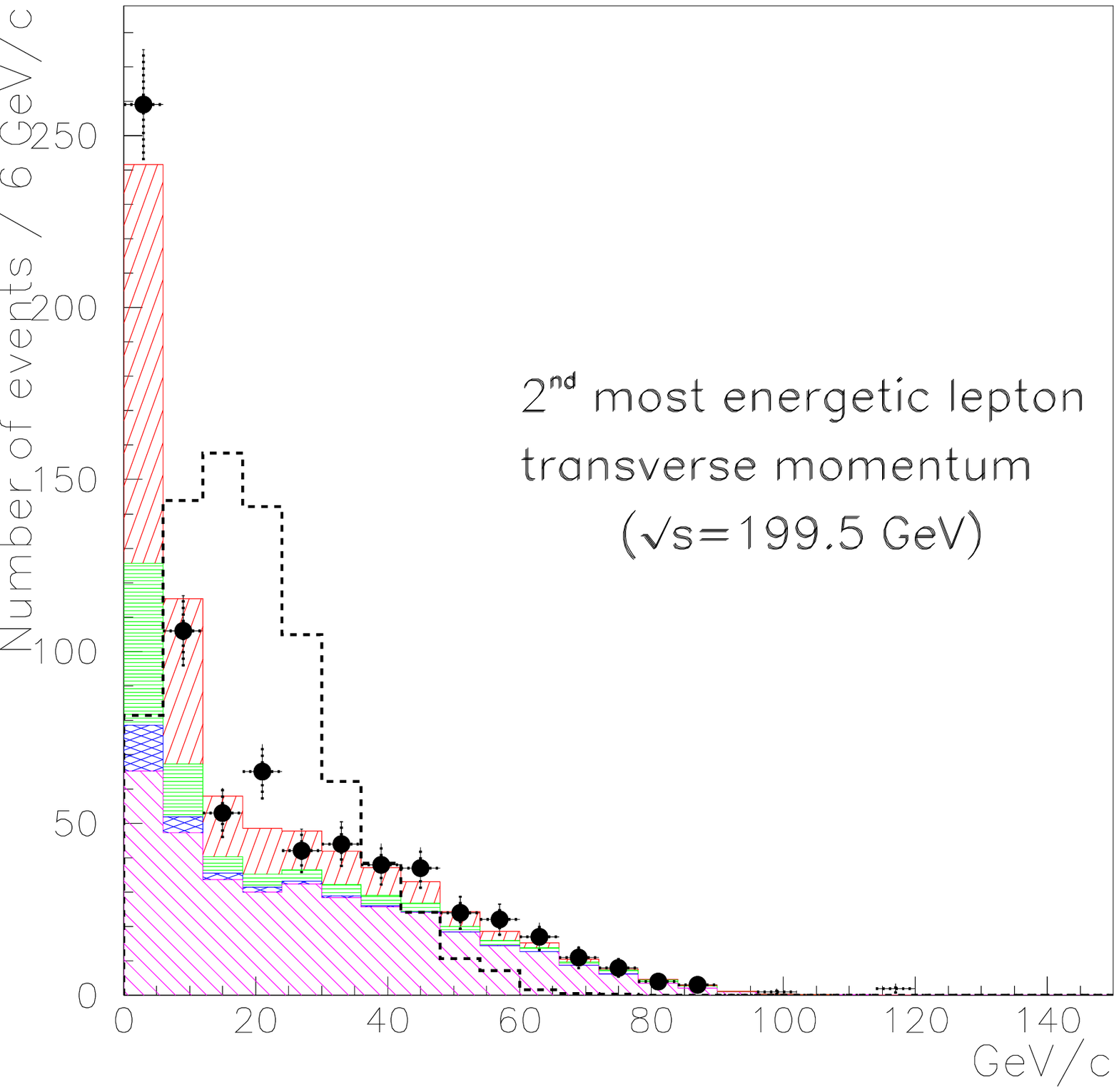,width=7.2cm}
\end{center}
\caption{Examples of data-simulation comparison at preselection level.
The dashed lines are examples of signal distributions (see text). 
%The legend for the other histograms shown on the first plot 
%is valid for the full page.
}
\label{fig:presel}
\end{figure}

\begin{figure}
\begin{center}
\epsfig{file=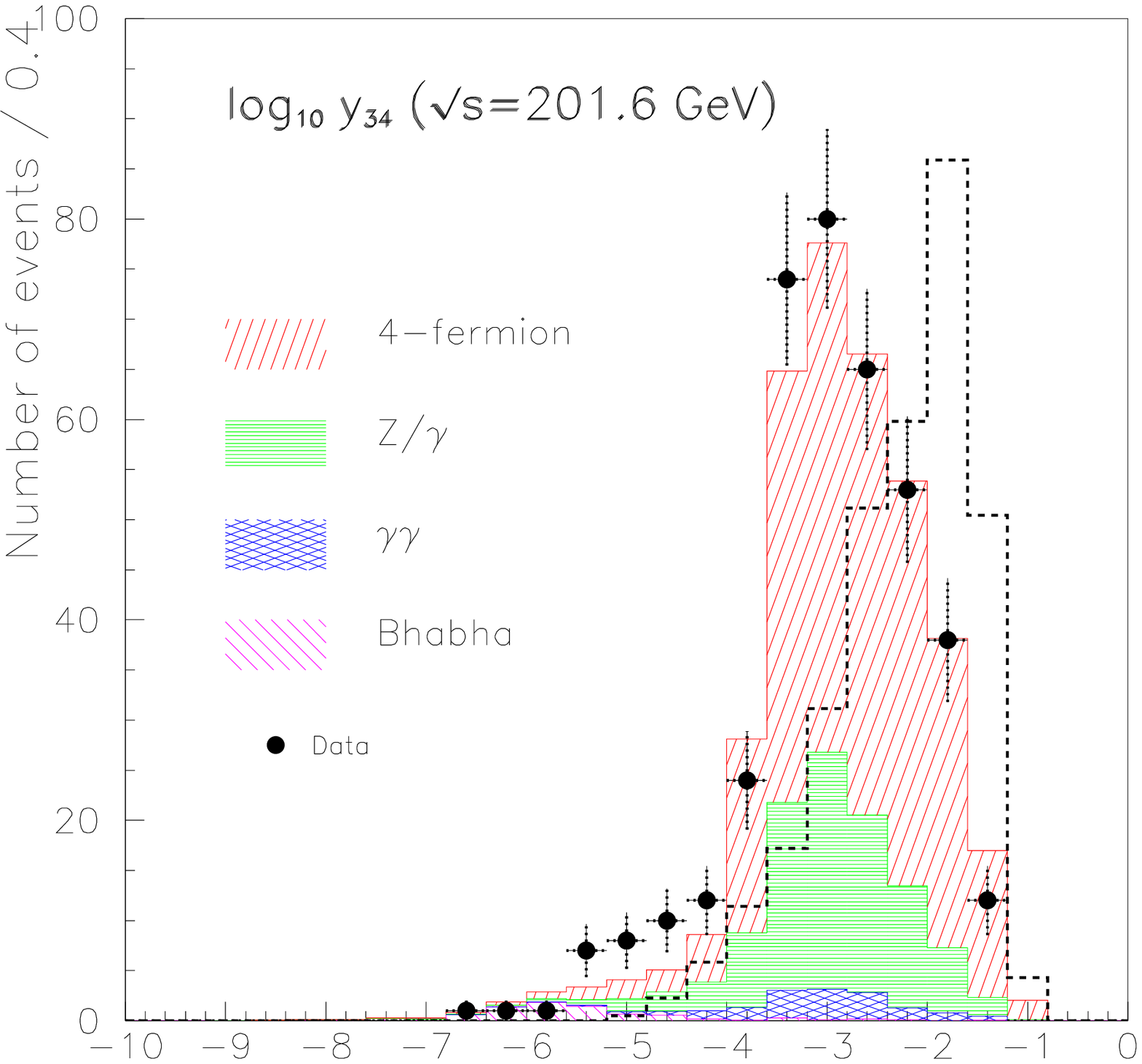,width=7.2cm}~\epsfig{file=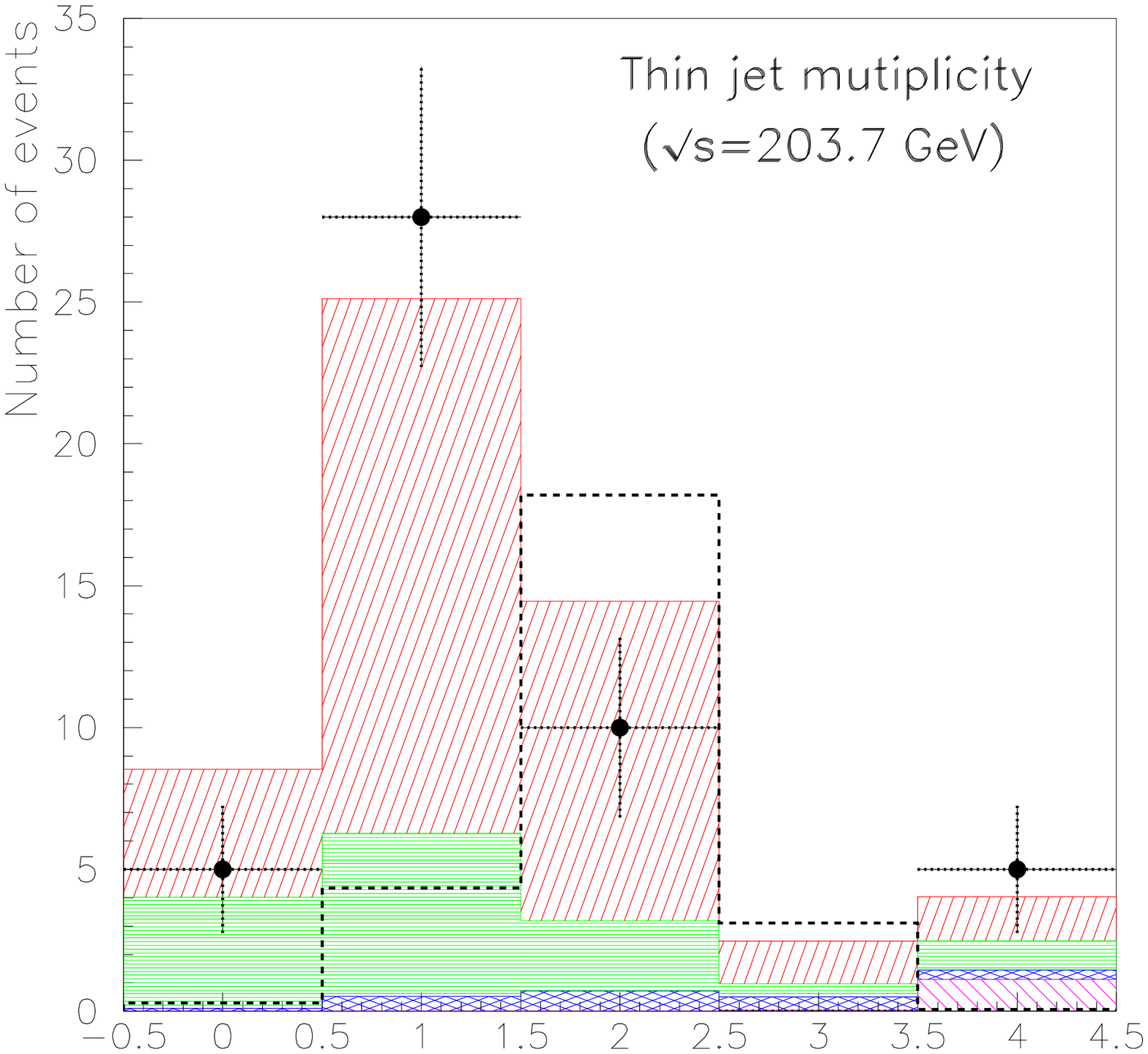,width=7.2cm}

\epsfig{file=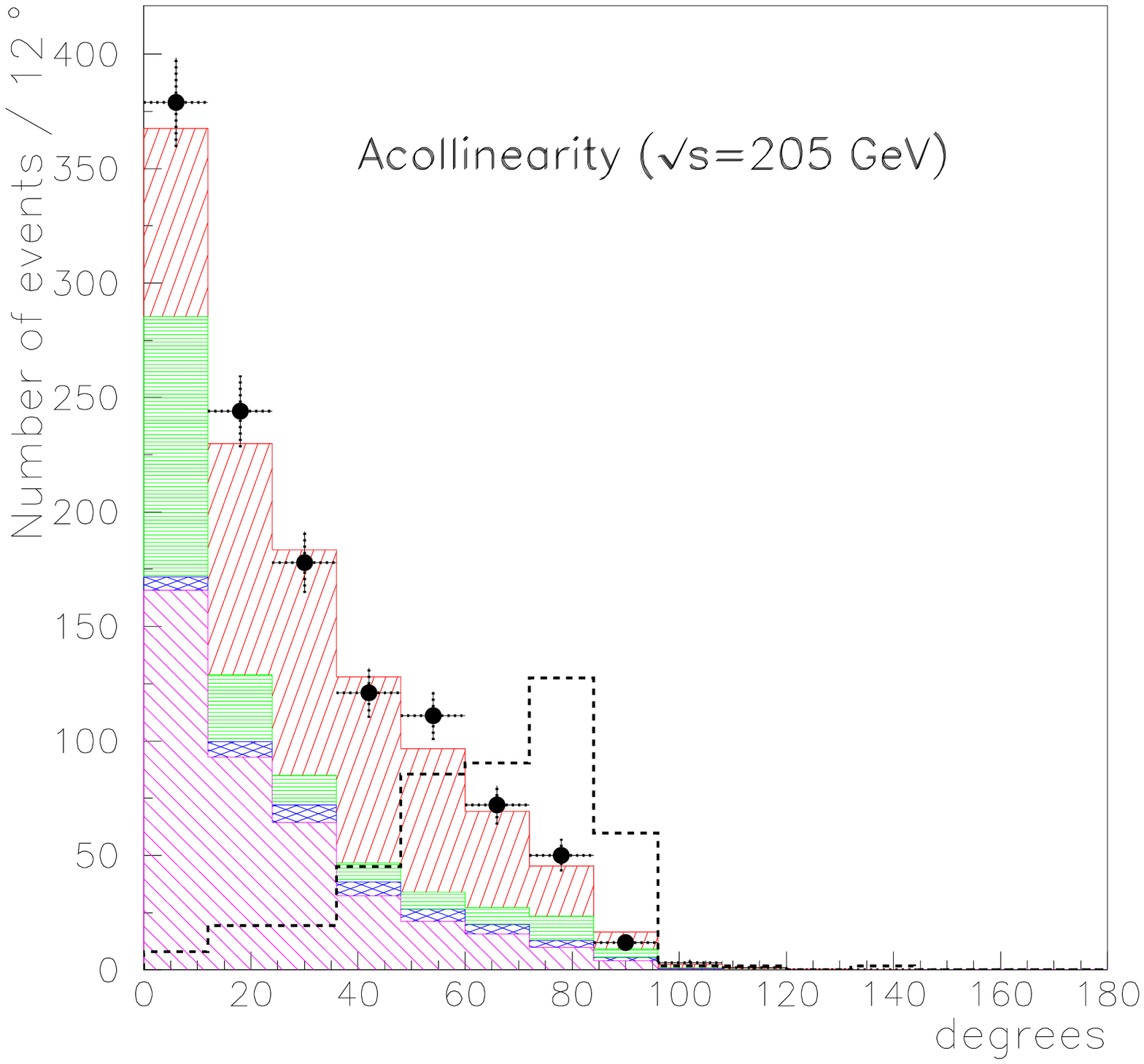,width=7.2cm}~\epsfig{file=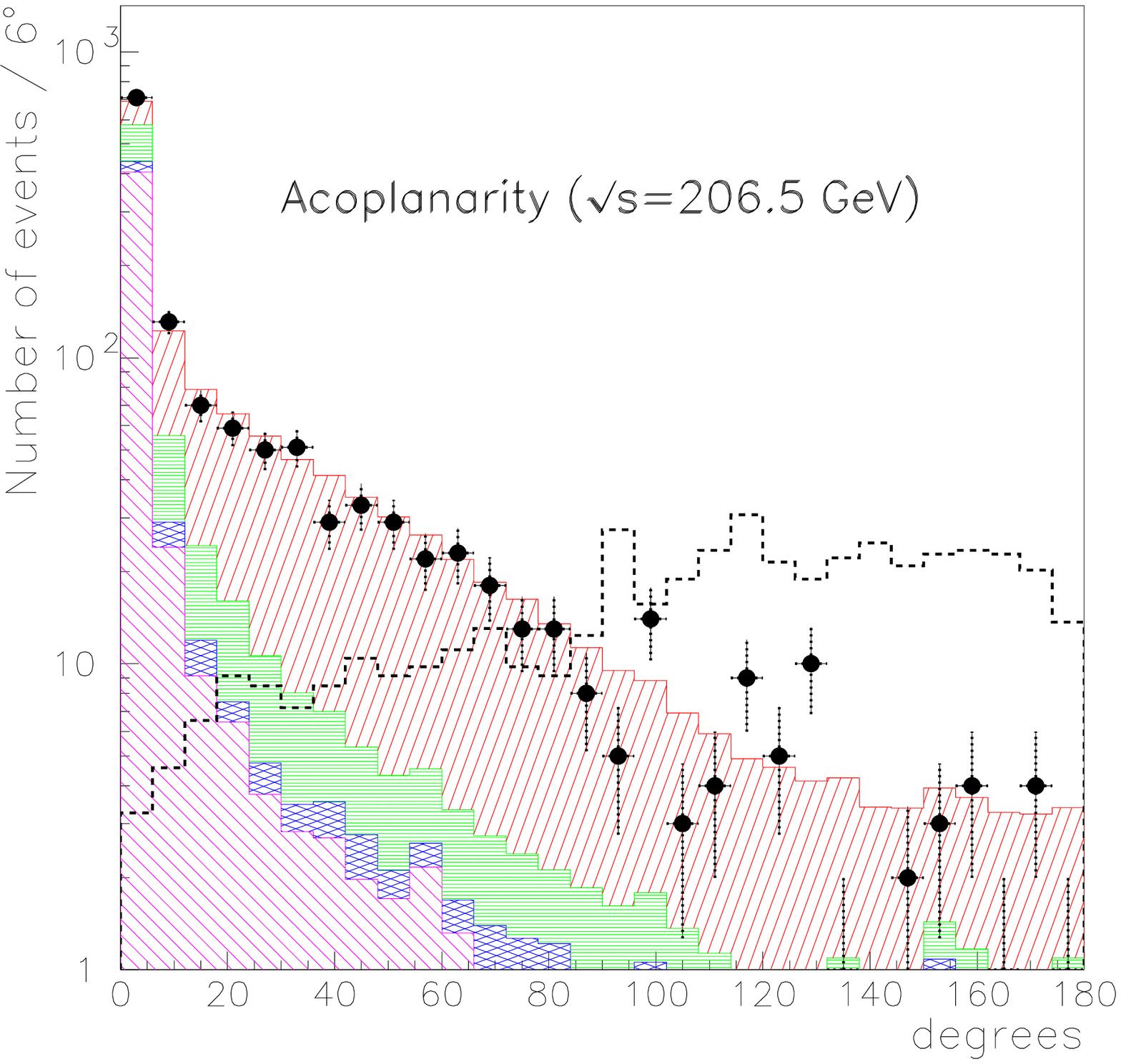,width=7.2cm}

\epsfig{file=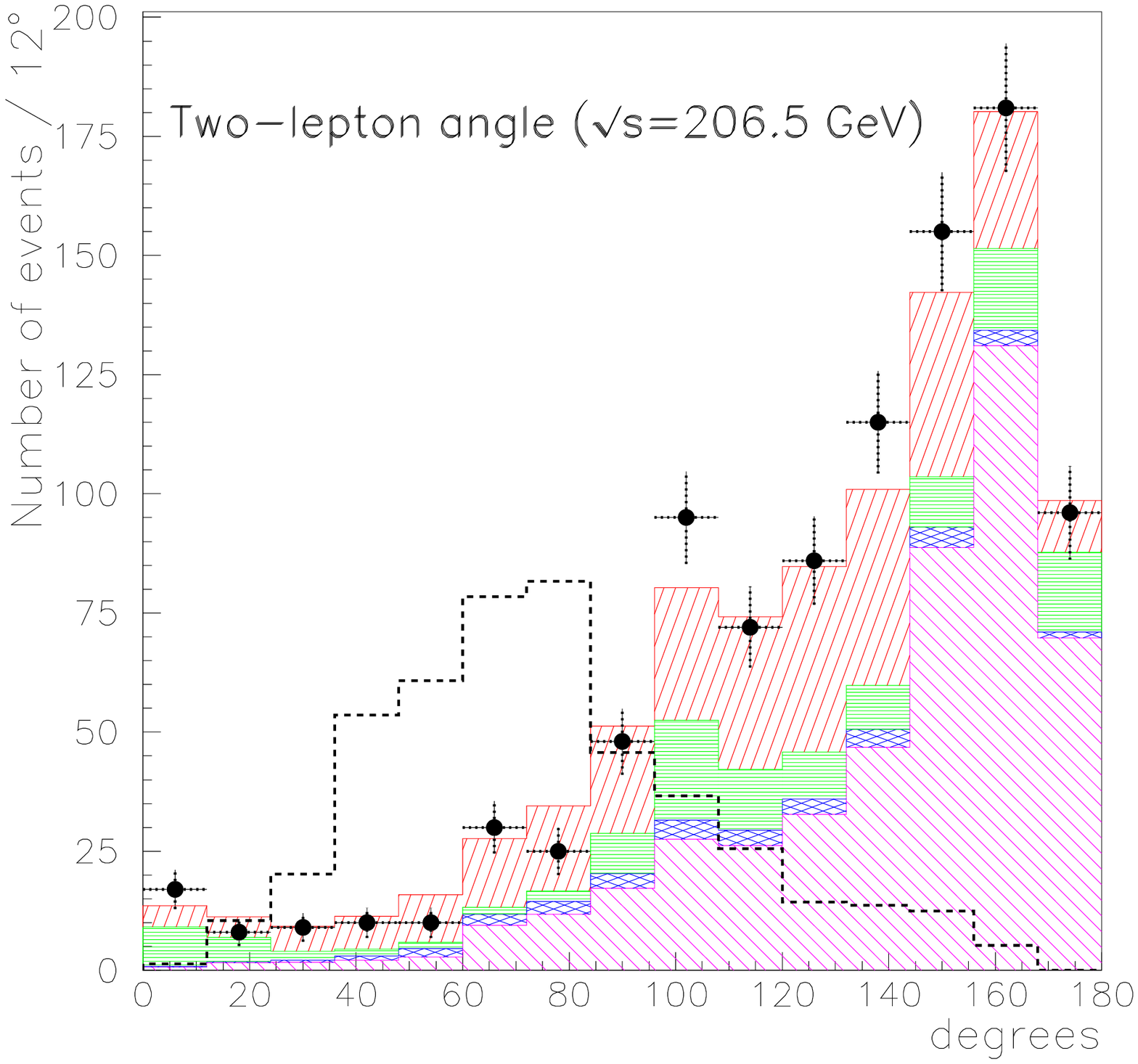,width=7.2cm}~\epsfig{file=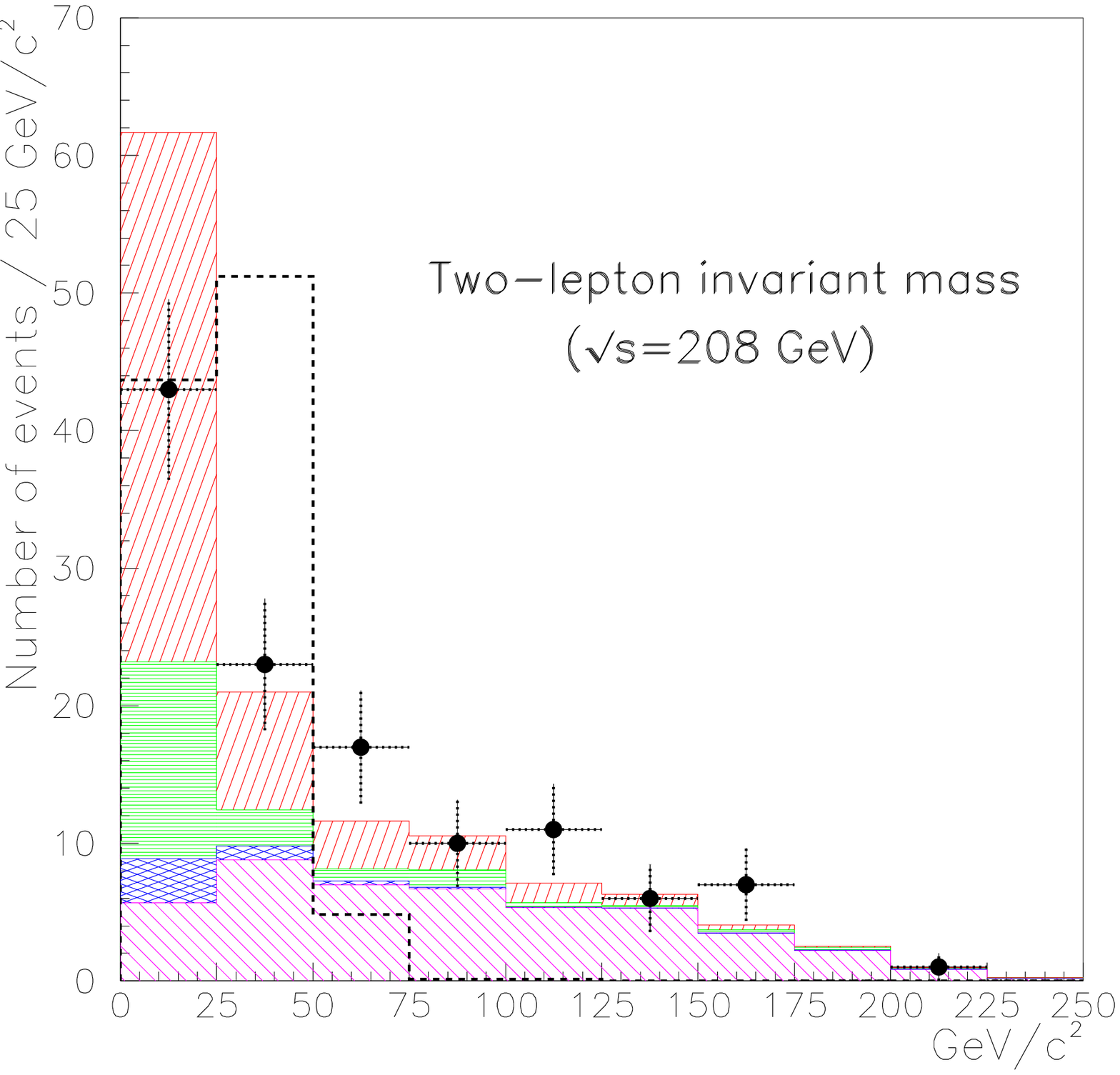,width=7.2cm}
\end{center}
\caption{Additional examples of data-simulation comparison at preselection level
(the $\log _{10}y_{34}$ and jet multiplicity distributions are given 
for events with at least four charged particles).
The dashed lines are examples of signal distributions (see text). 
%The legend for the other histograms shown on the first plot
%is valid for the full page.
}
\label{fig:presel2}
\end{figure}

The numbers
of data and of SM Monte Carlo events after the preselection
are shown in Table~\ref{tab:presel}. 
Distributions of some important variables
are shown in Figures~\ref{fig:presel} 
and~\ref{fig:presel2} at the
preselection level. The plots were chosen so as to
give at least an example of each centre-of-mass energy.
The agreement between real data and
simulated SM background is good.

Some examples of signal distributions are also given,
for the following parameters: 
$\lambda _{1j1}=0.05$,
$m_0=\sqrt{s}$,
$\tan{\beta}=1.5$,
$\mu=-125$~GeV/$c^2$,
$M_2=115$~GeV/$c^2$.
The missing energy signal is given for the semi-leptonic
channel, $j=3$, and scaled by a factor of 5;
the second lepton transverse momentum 
and the thin jet multiplicity signals are given for the semi-leptonic
channel, $j=2$, and scaled by a factor of 5 and 2 respectively;
the $\log _{10}y_{34}$ signal is given for the four-lepton
channel, $j=3$, and scaled by a factor of 20;
the acollinearity and the two-lepton angle signals 
are given for the two-lepton
channel, $j=2$, and scaled by a factor of 20;
the acoplanarity and the two-lepton invariant mass signals 
are given for the two-lepton channel, $j=3$, 
and scaled by a factor of 20 and 50 respectively.\\

\begin{figure}
\begin{center}
\epsfig{file=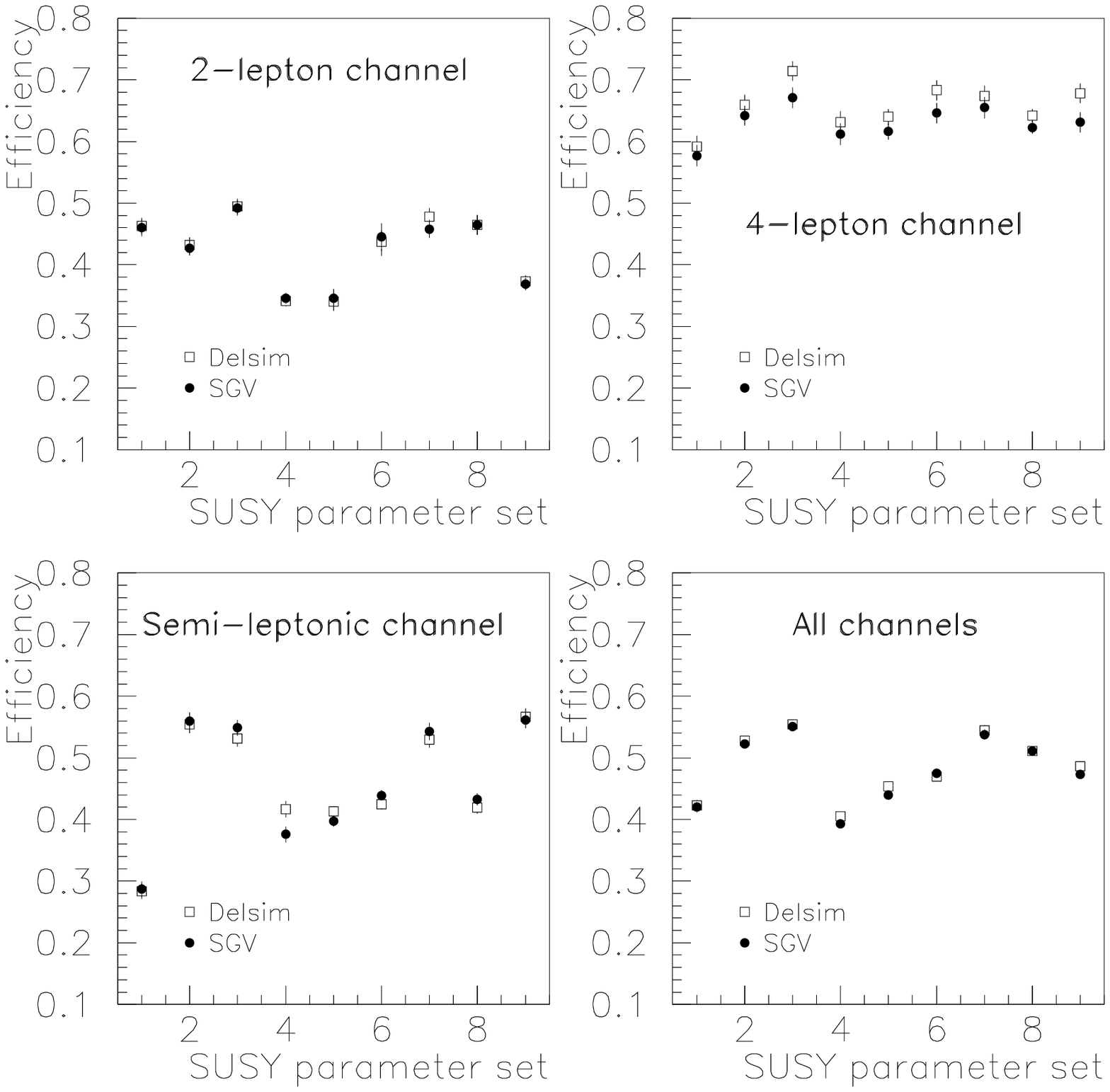,width=14cm}
\end{center}
\caption{{\tt SGV}-{\tt DELSIM} comparison for nine SUSY parameter sets 
with different sneutrino masses and total widths (see Table~\ref{tab:param})
in the case of a $\lambda _{121}$ coupling and for
$\tan{\beta}=1.5$, $\sqrt{s}=206.5$~GeV. For this comparison, the same
number of events was simulated with SGV and DELSIM.}
\label{fig:sgv}
\end{figure}

The efficiency of the selections, including the preselection,
depends on the SUSY parameters. 
It is shown in Figure~\ref{fig:sgv} for the nine parameter sets
which were fully simulated with {\tt DELSIM} (Table~\ref{tab:param})
using a $\lambda _{121}$ coupling and $\sqrt{s}=206.5$~GeV.
%is for example at $\sqrt{s}\simeq 200$~GeV
%for the parameter set
%($M_{\tilde\nu}=\sqrt{s}$, $\mu=-125$~GeV/$c^2$, $M_2=115$~GeV/$c^2$, 
%$\tan{\beta}=1.5$)
%$47\%$, $69\%$ and $44\%$ respectively on
%two-lepton, four or six lepton and semi-leptonic $\lambda_{121}$ signals
%and 37\%, 44\%, 21\% on $\lambda_{131}$ signals.
The Figure also shows the efficiencies evaluated with {\tt SGV};
they are compatible. The fact that {\tt SGV} efficiencies were
systematically lower than {\tt DELSIM} efficiencies in the 
four or six lepton channel was not compensated for, first because
this has a small impact on the global efficiency and second
because it gives conservative results.
The efficiencies for a $\lambda _{131}$ coupling are comparable, although
always lower.

The expected background at the end of the selections is mainly
composed of four-fermion events. The $\gamma\gamma$ background
is totally negligible. The background coming from the
two-fermion events is small and its proportion 
decreases when the centre-of-mass energy increases.
For instance, it is 9\% in the two-lepton channel
at 183~GeV and 4\% at 206.5~GeV.

\begin{table}[bt]
\begin{center}
\begin{tabular}{|c|rr|}
\hline
$\sqrt{s}$ (GeV) & SM~~~~~ & Data \\
\hline
182.7 & 1003$\pm$11 & 1071 \\
188.6 & 2909$\pm$16 & 2913 \\
191.6 &  466$\pm$3~\, & 484 \\
195.5 & 1387$\pm$8~\, & 1466 \\
199.5 & 1487$\pm$8~\, & 1537 \\
201.6 &  772$\pm$5~\, & 777 \\
203.7 &  110$\pm$1~\, & 99 \\
205.0 & 1180$\pm$7~\, & 1210 \\
206.5 & 1345$\pm$6~\, & 1322 \\
208.0 &  126$\pm$1~\, & 115 \\
\hline
\end{tabular}
\end{center}
\caption{Number of events after the preselection
(SM= Monte Carlo simulated SM background normalised to the data luminosity). 
Errors are statistical only.}
\label{tab:presel}
\end{table}

The number
of data and of SM Monte Carlo events at the end of the selections
is shown in Table~\ref{tab:res}. 
There is no significant excess of data in any of the three channels
and in any of the centre-of-mass energy samples.

\begin{table}[bt]
\begin{center}
\begin{tabular}{|ll||c|c|cc||cc|}
\hline
 &  & 2 leptons & 4 leptons & \multicolumn{2}{c||}{semi-leptonic} & 
 \multicolumn{2}{c|}{all channels} \\
$\sqrt{s}$ &(GeV)&         &           & $\lambda _{121}$& $\lambda _{131}$ 
 & $\lambda _{121}$& $\lambda _{131}$ \\
\hline
\hline
182.7 &SM    & 9.3$\pm$0.5 & 3.0$\pm$0.3 & 2.0$\pm$0.1 & 2.5$\pm$0.1 
& 14.3$\pm$0.6 & 14.8$\pm$0.6 \\
&Data & 7 & 2 & 4 & 0 &13 &9 \\
\hline
188.6 &SM    & 26.0$\pm$0.5 & 8.4$\pm$0.4 & 7.3$\pm$0.2 & 7.5$\pm$0.2 
&41.7$\pm$0.7 & 41.9$\pm$0.7\\
&Data & 26 & 7 & 8 & 2  &41 & 35\\
\hline
191.6 &SM    & 4.1$\pm$0.1 & 1.5$\pm$0.1 & 1.31$\pm$0.03 & 1.24$\pm$0.03 
&6.9$\pm$0.2 & 6.8$\pm$0.2\\
&Data & 2 & 2 & 2 & 2 & 6 & 6\\
\hline
195.5 &SM    & 11.9$\pm$0.2 & 3.9$\pm$0.2 & 4.2$\pm$0.1 & 3.7$\pm$0.1 
& 20.0$\pm$0.3 & 19.5$\pm$0.3\\
&Data & 10 & 5 & 5 & 2 & 20 & 17 \\
\hline
199.5 &SM    & 13.0$\pm$0.3 & 4.4$\pm$0.2 & 4.8$\pm$0.1 & 4.0$\pm$0.1 
& 22.2$\pm$0.4 & 21.4$\pm$0.4\\
&Data & 11 & 4 & 5 & 2  & 20 & 17\\
\hline
201.6 &SM    & 6.3$\pm$0.1 & 2.2$\pm$0.1 & 2.5$\pm$0.1 & 1.9$\pm$0.1 
& 11.0$\pm$0.2 & 10.4$\pm$0.2\\
&Data & 3 & 1 & 3 & 3 & 7 & 7\\
\hline
203.7 &SM    & 0.93$\pm$0.03 & 0.34$\pm$0.02 & 0.39$\pm$0.02 & 0.30$\pm$0.02 
&1.6$\pm$0.1 & 1.5$\pm$0.1\\
&Data & 0 & 0 & 0 & 0 & 0 & 0\\
\hline
205.0 &SM    & 10.1$\pm$0.3 & 3.5$\pm$0.2 & 3.9$\pm$0.2 & 3.2$\pm$0.2 
& 17.5$\pm$0.4 & 16.8$\pm$0.4\\
&Data & 14 & 4 & 4 & 3 &22 & 21 \\
\hline
206.5 &SM    & 11.4$\pm$0.3 & 4.4$\pm$0.2 & 4.7$\pm$0.3 & 3.6$\pm$0.3 
&20.5$\pm$0.5 & 19.4$\pm$0.5\\
&Data & 12 & 1 & 7 & 3 &20 & 16 \\
\hline
208.0 &SM    & 1.04$\pm$0.02 & 0.44$\pm$0.04 & 0.50$\pm$0.03 & 0.36$\pm$0.03 
&1.9$\pm$0.1 & 1.8$\pm$0.1\\
&Data & 1 & 0 & 0 & 0 & 1 & 1\\
\hline
\end{tabular}
\end{center}
\caption{Number of events at the end of the selection
(SM = Monte Carlo simulated SM background). Errors are statistical only.
Note that the sum of all channels is not independent for
$\lambda _{121}$ and $\lambda _{131}$ due to the
common analysis for the first two channels.}
\label{tab:res}
\end{table}

\section{Limits on $\lambda _{121}$ and $\lambda _{131}$ couplings}

Besides being used as an event generator,
{\tt SUSYGEN} was also used to scan a wide part
of the MSSM parameter space
and compute all the cross-sections of the signal,
with $\lambda _{1j1}=5\times 10^{-3}$.
In the model adopted for this search and described
in section~1, all SUSY phenomenology can be derived
from the four parameters $\tan{\beta}$, $m_0$, $M_2$ and $\mu$,
plus the centre-of-mass energy for the kinematics.
The parameter sets explored in the scan were:
\begin{itemize}
\item $\sqrt{s}= 182.7$, 188.6, 199.5 and 206.5~GeV,
\item $\tan{\beta}=1.5$ or 30,
\item $m_0$= 100 to 230 GeV/$c^2$
      (170 to 215~GeV/$c^2$ in steps of 1~GeV/$c^2$,
       100 to 170~GeV/$c^2$ in steps of 10~GeV/$c^2$,
       215 to 230~GeV/$c^2$ in steps of 5~GeV/$c^2$),
\item $M_2= 5$ to 405~GeV/$c^2$ in steps of 10~GeV/$c^2$,
\item $\mu = -305$ to $305$~GeV/$c^2$ in steps of 10~GeV/$c^2$.
\end{itemize}

The cross-sections at $\sqrt{s}=191.6$~GeV
were taken from the $\sqrt{s}=188.6$~GeV scan simply
assuming $\sqrt{s}+3$~GeV;
in the same way,
the cross-sections at 195.5 and 201.6~GeV
were taken from the 199.5~GeV scan
and the cross-sections at 203.7, 205.0 and 208.0~GeV
were taken from the 206.5~GeV scan
assuming the corresponding centre-of-mass energy shifts. 

The sneutrino mass $M_{\tilde\nu}$ was assigned the value of $m_0$,
thus slightly departing from a strict mSUGRA model.
%for reasons of easiness. 
This is a conservative hypothesis since the sneutrino mass
tends to decrease whereas the gaugino masses are left untouched.

Very small values of the sneutrino total width 
($\Gamma _{\tilde\nu}<150$~MeV/$c^2$)
correspond to regions of the parameter
space where the sneutrino is lighter than the gauginos;
they can hardly be detected by the present analysis, 
however they are covered by $e^+e^-\to l^+l^-$ analyses~\cite{direct}.

Regions of the parameter space already excluded
by the precision measurements at LEP1 were not 
further explored in the present scans.
This condition was implemented in the following way.
Using the expression for the cross-section
at the resonance, $\sigma(e^+e^-\to Z\to X)=
12\pi\frac{\Gamma _{ee}\Gamma _X}{M_Z^2\Gamma _Z^2}$,
%(under the assumption that the unpredicted decay modes
%of the $Z$ are pair production  via the $s$-channel)
the limit $\Gamma ^{\mbox{\small new}}<6.6$~MeV/$c^2$
%% LEP combined: $\Gamma ^{\mbox{\small new}}<6.6$~MeV 
at 95\%~CL~\cite{lep1} can be converted into
an upper limit on the cross-section of new decay modes:
$\sigma ^{\mbox{new}}<157.2$~pb at 95\%~CL.
%% LEP combined: $\sigma ^{\mbox{new}}<47.6$~pb.
SUSY parameter sets for which the total cross-section
of pair production of charginos and neutralinos
%(which are all visible) 
at $\sqrt{s}\simeq M_Z$
was larger than this limit
were considered as excluded by LEP1.\\
%Since the $t$-channel contribution significantly affects
%the cross-section only for low values of $m_0$
%and only when the $s$-channel alone cross-section is small
%(much below 157~pb), the $s$-channel assumption
%is not expected to bias the result.\\

To derive the limits on one $\lambda$ coupling, 
each centre-of-mass energy was first considered separately.
The three channels being totally independent due to the
charged particle multiplicity criterion, 
they were summed up (Table~\ref{tab:res}).
On the other hand,
the ($M_{\tilde\nu} ,\Gamma _{\tilde\nu}$) plane was divided
into rectangular bins of size (1~GeV/$c^2$, 50~MeV/$c^2$).
For each set of parameters entering a given bin, 
the output of two scans was used.
First the {\tt SUSYGEN} scan to get $\sigma$,
the total cross-section
expected for $e^+e^-\to\tilde\nu\to X$, $X$ representing 
all final states from indirect sneutrino decays.
From the expression %for the cross-section 
given in the introduction,
$\sigma =\sigma _0\times\lambda ^2$. 
$\sigma _0$ is almost independent of $\lambda$ as long as 
$\lambda$ is reasonably small: for $\lambda=0.1$, $\sigma _0$
is at most a few percent lower than for $\lambda=10^{-4}$.
%\footnote{For instance, for $\tan{\beta}=5.$,
%$m_0=190$~GeV/$c^2$, $\mu=-150$~GeV/$c^2$ and $M_2=100$~GeV/$c^2$,
%% p. 20 de mon cahier
%$\sigma _0(\lambda =0.1)=0.975\times \sigma _0(\lambda =10^{-4})$.}.
Second a {\tt SGV} scan on the same parameters 
(however with wider steps, 
of 10~GeV on the full $m_0$ range and of 20~GeV for $\mu$ and $M_2$), 
to obtain the global
efficiency of the analysis, $\epsilon$, combining the three separate
channel efficiencies according to the branching
ratios predicted by {\tt SUSYGEN}. 1000 events were simulated
for each parameter set.
% on a verifie qu'il ne se passe rien de special pour
% les efficacites autour de la resonance, voir cahier p. 49

An upper limit at $95\%$~CL on the total
number of signal events ($N_{up}$) 
compatible with the data and expected background
was then calculated for each ($M_{\tilde\nu} ,\Gamma _{\tilde\nu}$) bin, 
combining the ten centre-of-mass
energy samples considered as independent samples. 
This was done using the Bayesian method
described in reference~\cite{vo}; the relative probabilities
of each centre-of-mass energy were taken as
$$ w_i = \frac{(\sigma _0\epsilon)_i {\int\!\cal{L}}_i}
  {{\displaystyle \sum _{j=1}^{10}} (\sigma _0\epsilon)_j {\int\!\cal{L}}_j},$$
where $(\sigma _0\epsilon)_i$ is the lowest value
of such a product in the considered bin
and $\int\!{\cal L}_i$ is the integrated luminosity
of the $i^{\rm th}$ sample.
The 95\%~CL upper limit on $\lambda$ in each bin
was then
$$ \lambda  < \sqrt{\frac{N_{up}}
{{\displaystyle \sum _{j=1}^{10}} (\sigma _0\epsilon)_j {\int\!\cal{L}}_j}}.$$
The whole procedure was repeated for the second coupling.\\

\begin{figure}[tbh]
\epsfig{file=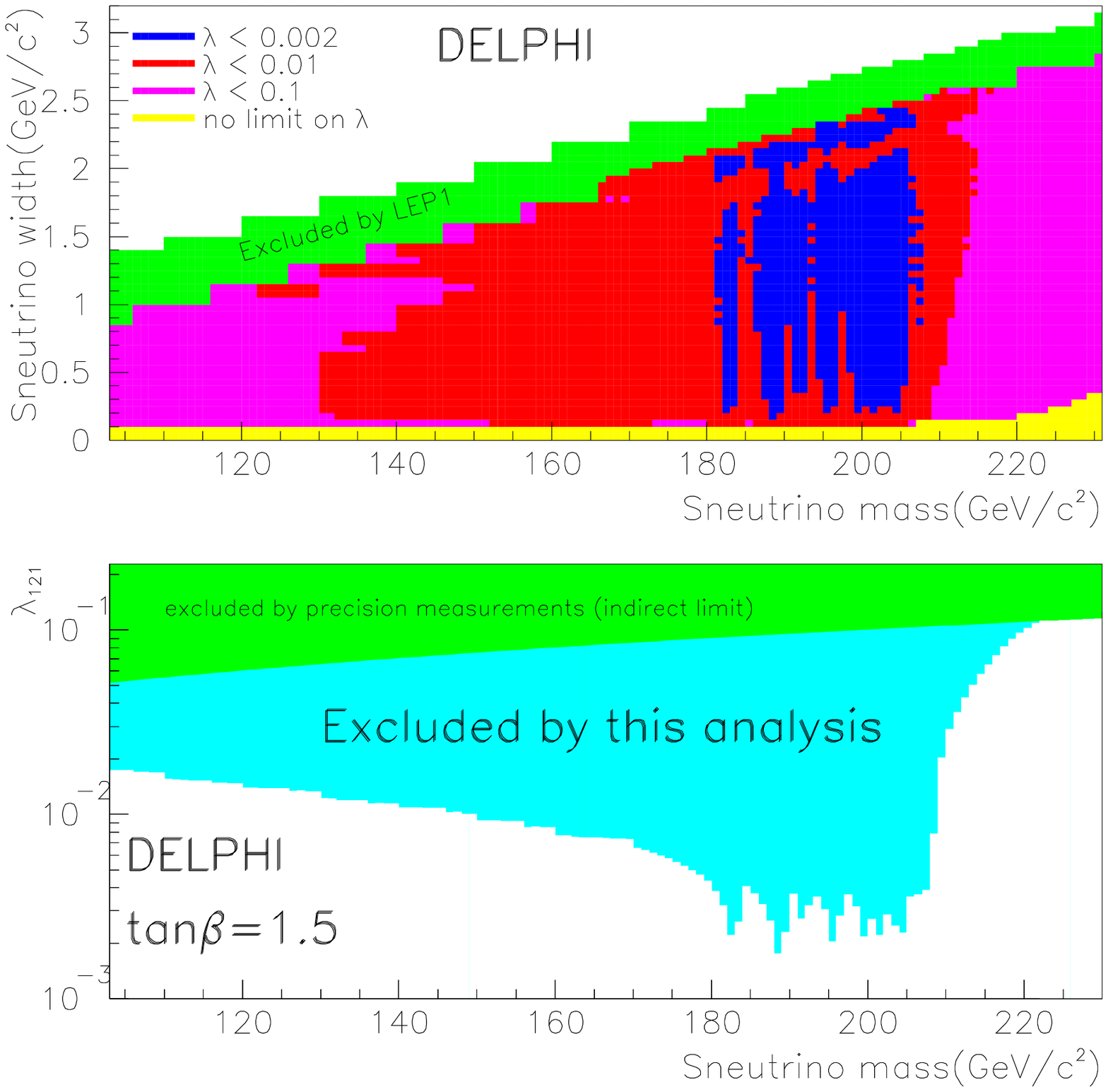,height=18.0cm}
\caption{For $\tan\beta =1.5$, upper limit on $\lambda _{121}$
as a function of $M_{\tilde\nu}$ and $\Gamma _{\tilde\nu}$ (top)
and as a function of $M_{\tilde\nu}$ 
assuming \mbox{$\Gamma_{\tilde\nu}>150$~MeV/$c^2$} 
(bottom).
The white zone in the top plot corresponds to non existing sneutrino widths
given the $\mu$ parameter range. 
The area entitled `no limit on $\lambda$' corresponds to upper limits
larger than 0.1.
The indirect limit coming from precision measurements is drawn
in the bottom plot assuming $M_{\tilde e_R}=M_{\tilde\nu}$.}
\label{fig:resulow2}
\end{figure}

\begin{figure}[tbh]
\epsfig{file=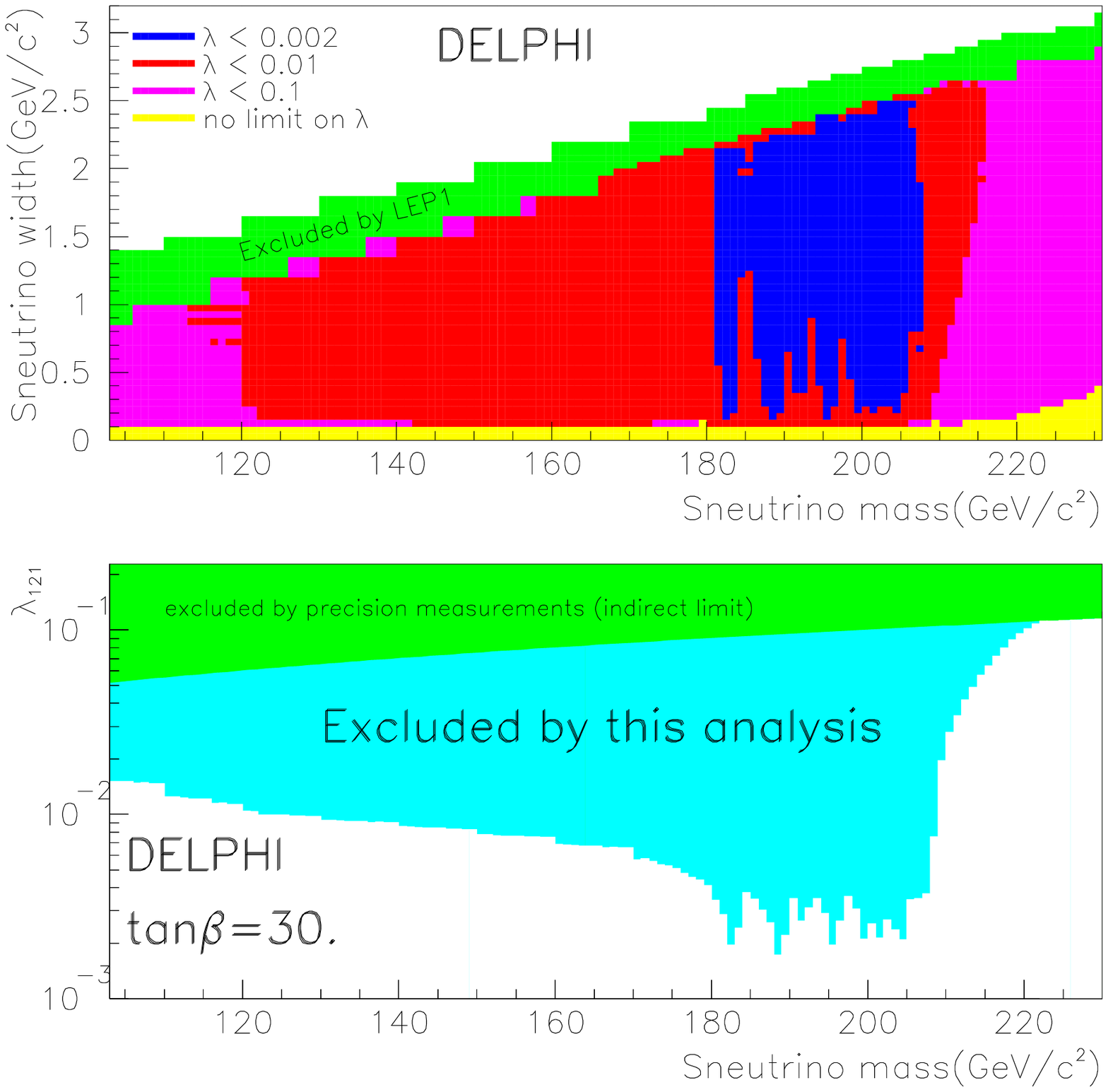,height=18.0cm}
\caption{For $\tan\beta =30$, upper limit on $\lambda _{121}$
as a function of $M_{\tilde\nu}$ and $\Gamma _{\tilde\nu}$ (top)
and as a function of $M_{\tilde\nu}$ 
assuming \mbox{$\Gamma_{\tilde\nu}>150$~MeV/$c^2$} 
(bottom).
The white zone in the top plot corresponds to non existing sneutrino widths
given the $\mu$ parameter range.
The area entitled `no limit on $\lambda$' corresponds to upper limits
larger than 0.1.
The indirect limit coming from precision measurements is drawn
in the bottom plot assuming $M_{\tilde e_R}=M_{\tilde\nu}$.}
\label{fig:resuhigh2}
\end{figure}

\begin{figure}[tbh]
\epsfig{file=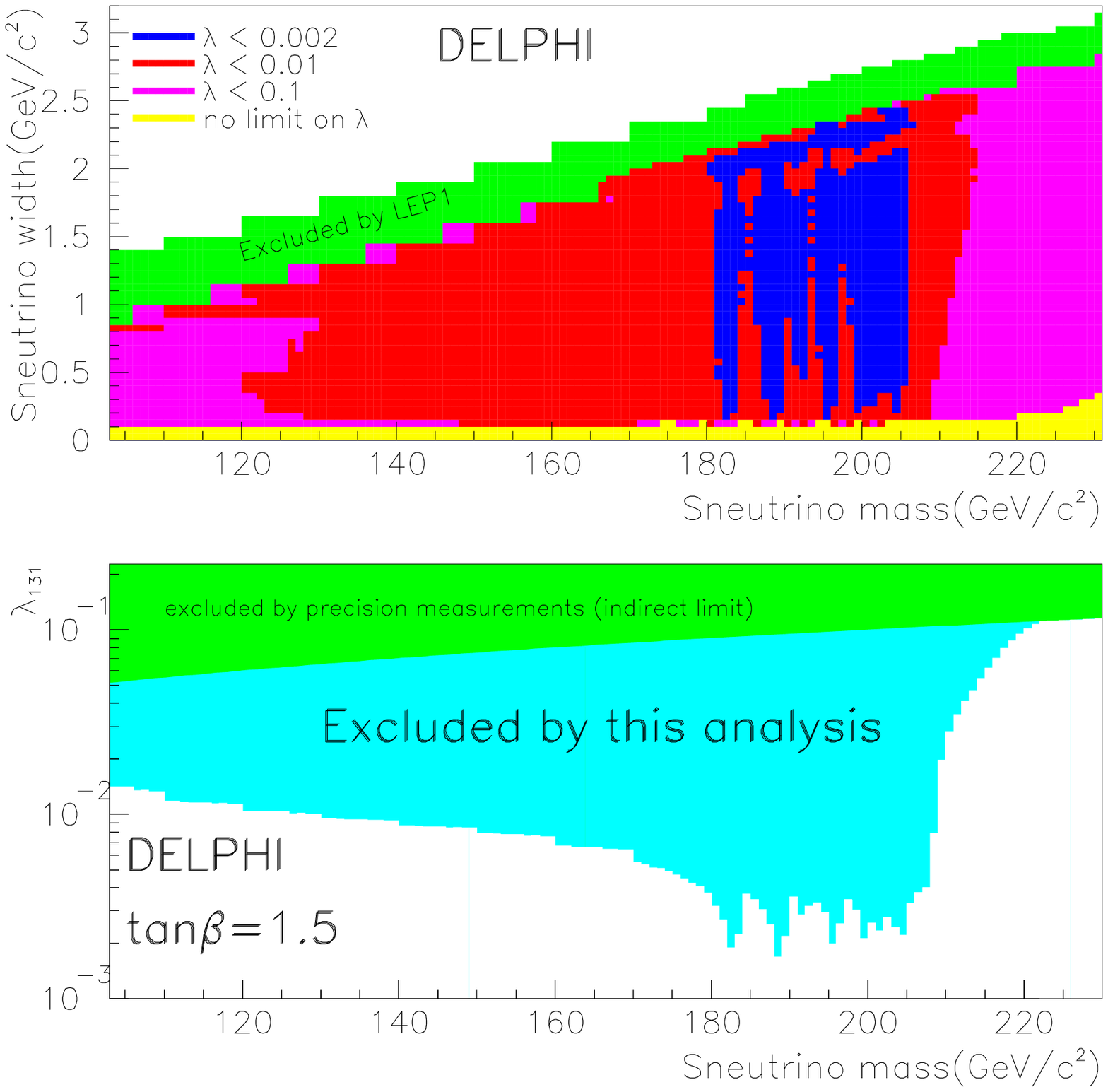,height=18.0cm}
\caption{For $\tan\beta =1.5$, upper limit on $\lambda _{131}$
as a function of $M_{\tilde\nu}$ and $\Gamma _{\tilde\nu}$ (top)
and as a function of $M_{\tilde\nu}$ 
assuming \mbox{$\Gamma_{\tilde\nu}>150$~MeV/$c^2$} 
(bottom).
The white zone in the top plot corresponds to non existing sneutrino widths
given the $\mu$ parameter range.
The area entitled `no limit on $\lambda$' corresponds to upper limits
larger than 0.1.
The indirect limit coming from precision measurements is drawn
in the bottom plot assuming $M_{\tilde e_R}=M_{\tilde\nu}$.}
\label{fig:resulow3}
\end{figure}

\begin{figure}[tbh]
\epsfig{file=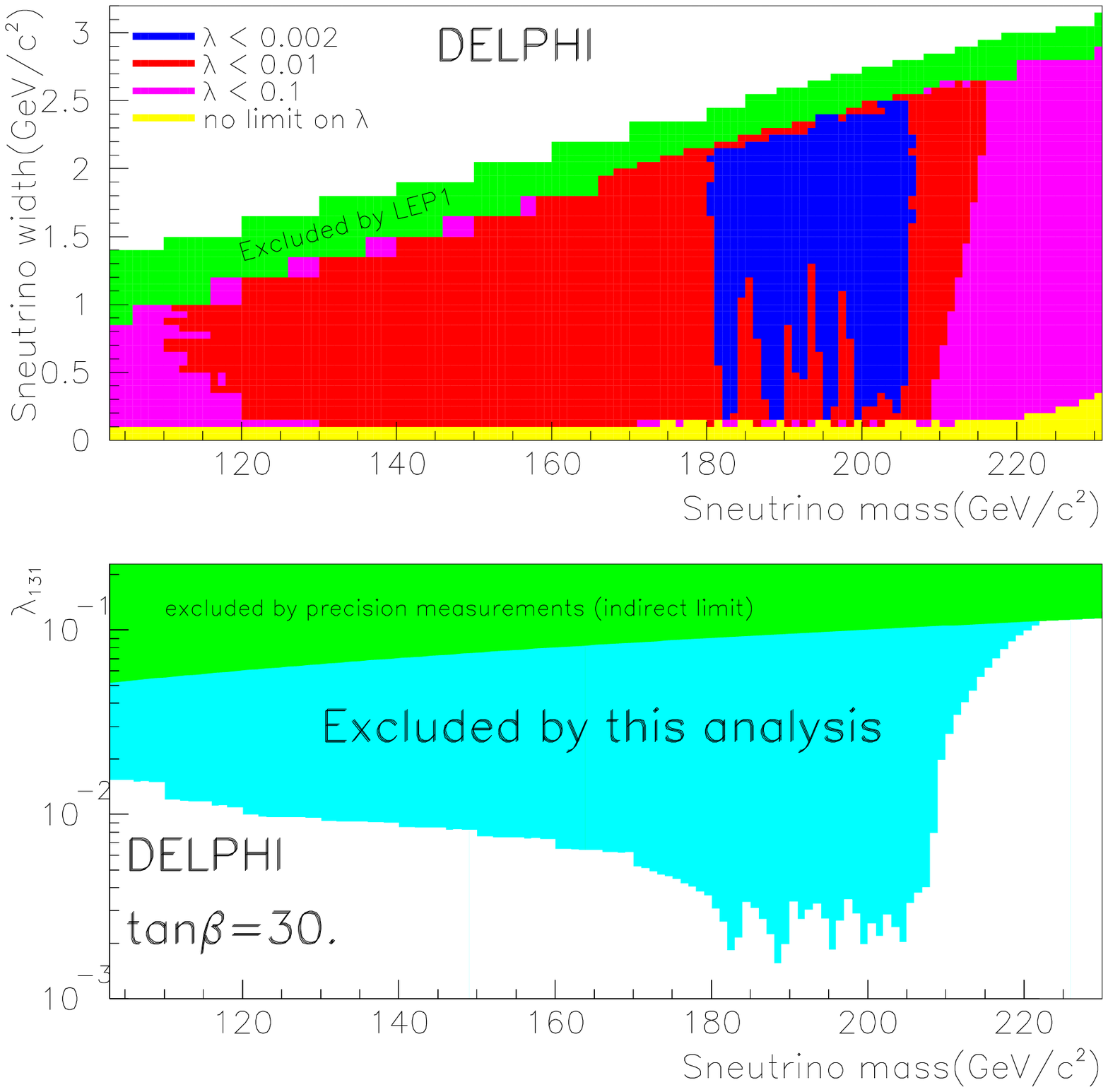,height=18.0cm}
\caption{For $\tan\beta =30$, upper limit on $\lambda _{131}$
as a function of $M_{\tilde\nu}$ and $\Gamma _{\tilde\nu}$ (top)
and as a function of $M_{\tilde\nu}$ 
assuming \mbox{$\Gamma_{\tilde\nu}>150$~MeV/$c^2$} 
(bottom).
The white zone in the top plot corresponds to non existing sneutrino widths
given the $\mu$ parameter range.
The area entitled `no limit on $\lambda$' corresponds to upper limits
larger than 0.1.
The indirect limit coming from precision measurements is drawn
in the bottom plot assuming $M_{\tilde e_R}=M_{\tilde\nu}$.}
\label{fig:resuhigh3}
\end{figure}

The results are shown in Figures \ref{fig:resulow2} to \ref{fig:resuhigh3}
corresponding to the two values of $\tan{\beta}$ considered.
In a very large fraction of the ($M_{\tilde\nu} ,\Gamma _{\tilde\nu}$) plane 
the obtained upper limit on $\lambda _{1j1}$ is at most 0.1, and
below 0.01 in still the major part of the area allowed by the parameters range.
%To enlarge the area tested in the  
%high $\Gamma _{\tilde\nu}$
%direction, one should enlarge the $\mu$ parameter range.
% voir mon cahier pp. 8 et 41
%Indeed, for large $|\mu |$, the LEP1 cross-section decreases
%and large values of $\Gamma _{\tilde\nu}$ appear that are
%not yet excluded.

There are two main sources of systematic errors on these results.
One is the estimation of the expected SM background. Uncertainties arise 
from the Monte Carlo statistics (at most $\pm$5\%, see Table~\ref{tab:res}),
from the detector response simulation
and from the cross-sections evaluation and event modelling.
In the last case, a comparison of different generators 
gave at most $\pm$5\% difference, mainly coming from the hadronisation
modelling, for the four-fermion processes which are
the dominant background. This uncertainty is smaller
for two-fermion processes and larger
for $\gamma\gamma$ events,
which however do not affect the final results because very few
events of that kind remain at the end of the selection.
%which are nevertheless negligible in the end.
The resulting systematic uncertainty on the limit on $\lambda$, 
evaluated separately for each centre-of-mass energy, 
%depends on the comparison between data and SM model predictions
is of the order of $\pm$2\%.
The second source of systematic errors 
is the estimation of the selection efficiency,
performed with a fast simulation and however statistically limited
(up to $\pm$5\% on low branching ratio channels). Here an important
uncertainty arises from the track reconstruction and
the lepton identification efficiencies. Discrepancies between {\tt DELSIM}
and {\tt SGV} are at the level of $\pm$4\% for the global efficiency
(Figure~\ref{fig:sgv}). The resulting variation of the limit
on $\lambda$, using the prescription of reference \cite{cousins},
is $\pm$3\% for the sample with highest luminosity,
and at most $\pm$1\% for the other samples.
%If the two sources are considered as independent and combined,
%the resulting relative uncertainty on $\lambda ^2$ is 6\%,
%and twice less on  $\lambda$. 
These effects were considered small and were not included
in the plots of Figures \ref{fig:resulow2} to \ref{fig:resuhigh3}.\\

In order to make the results easier to read, 
the upper limits on the $\lambda$ couplings were also derived as a function
of the sneutrino mass only. This was simply done by keeping
the most conservative limit for each sneutrino mass,
assuming $\Gamma _{\tilde\nu}>150$~MeV/$c^2$.
The results can be seen in the same Figures as for
the two-dimensional limits. The limits clearly show
the centre-of-mass energy structure of the data samples;
they are especially stringent for $M_{\tilde\nu}\simeq\sqrt{s}$.

\section{Conclusion}

The possibility of single production of supersymmetric
particles was explored
but none was seen.
%No new resonance has been observed.
Upper limits at the 95\%~CL were
derived in the mSUGRA constrained MSSM framework
with low and high values of $\tan{\beta}$
for the two possible
$R_p$ violating couplings $\lambda_{121}$ and $\lambda_{131}$.
They are at the level of 2 to 3$\times 10^{-3}$, depending on $M_{\tilde\nu}$,
when it is close to the centre-of-mass energy.
They are slightly better for high $\tan{\beta}$ and slightly
worse for $\lambda _{131}$ as compared to $\lambda _{121}$.
In any case, they are one or two orders of magnitude better
than the published indirect limit, even for sneutrino masses
outside the centre-of-mass energies of the analysed data, 
due to the $t$-channel %above the resonance 
and, more importantly, to the initial state radiation
below the resonance.

%         Modified on 04-06-1999 by dimartino
%-------------------------------------------------------------------
\subsection*{Acknowledgements}
\vskip 3 mm
 We are greatly indebted to our technical 
collaborators, to the members of the CERN-SL Division for the excellent 
performance of the LEP collider, and to the funding agencies for their
support in building and operating the DELPHI detector.\\
We acknowledge in particular the support of \\
Austrian Federal Ministry of Education, Science and Culture,
GZ 616.364/2-III/2a/98, \\
FNRS--FWO, Flanders Institute to encourage scientific and technological 
research in the industry (IWT), Belgium,  \\
FINEP, CNPq, CAPES, FUJB and FAPERJ, Brazil, \\
Czech Ministry of Industry and Trade, GA CR 202/99/1362,\\
Commission of the European Communities (DG XII), \\
Direction des Sciences de la Mati$\grave{\mbox{\rm e}}$re, CEA, France, \\
Bundesministerium f$\ddot{\mbox{\rm u}}$r Bildung, Wissenschaft, Forschung 
und Technologie, Germany,\\
General Secretariat for Research and Technology, Greece, \\
National Science Foundation (NWO) and Foundation for Research on Matter (FOM),
The Netherlands, \\
Norwegian Research Council,  \\
State Committee for Scientific Research, Poland, SPUB-M/CERN/PO3/DZ296/2000,
SPUB-M/CERN/PO3/DZ297/2000, 2P03B 104 19 and 2P03B 69 23(2002-2004)\\
JNICT--Junta Nacional de Investiga\c{c}\~{a}o Cient\'{\i}fica 
e Tecnol$\acute{\mbox{\rm o}}$gica, Portugal, \\
Vedecka grantova agentura MS SR, Slovakia, Nr. 95/5195/134, \\
Ministry of Science and Technology of the Republic of Slovenia, \\
CICYT, Spain, AEN99-0950 and AEN99-0761,  \\
The Swedish Natural Science Research Council,      \\
Particle Physics and Astronomy Research Council, UK, \\
Department of Energy, USA, DE-FG02-01ER41155. 
%=========================================================================%


\begin{thebibliography}{99}
\bibitem{mssm} H.P. Nilles, Phys. Rep. {\bf 110} (1984) 1;\\
               H.E. Haber and G.L. Kane, Phys. Rep. {\bf 117} (1985) 75.
\bibitem{fayet} P. Fayet, Phys. Lett. {\bf B69} (1977) 489;\\
                G. Farrar and P. Fayet, Phys. Lett. {\bf B76} (1978) 575.
\bibitem{hall} S. Dimopoulos and L.J. Hall, 
               Phys. Lett. {\bf B207} (1988) 210.
\bibitem{barger} V. Barger et al., Phys. Rev. {\bf D40} (1989) 2987.
\bibitem{susygen} S. Katsanevas and P. Morawitz,
                  Comput. Phys. Commun. {\bf 112} (1998) 227.
\bibitem{indir} H. Dreiner, in {\it Perspectives on Supersymmetry},
                edited by G.L. Kane (World Scientific, Singapore, 1998)
                p. 462.
\bibitem{direct} R. Barate et al., ALEPH coll.,
                 Eur. Phys. J. {\bf C12} (2000) 183;\\
                 P. Abreu et al., DELPHI coll.,
                 Phys. Lett. {\bf B485} (2000) 45;\\
                 M. Acciarri et al., L3 coll.,
                 Phys. Lett. {\bf B489} (2000) 81;\\
                 G. Abbiendi et al., OPAL coll.,
                 Eur. Phys. J. {\bf C13} (2000) 553.
%\bibitem{corinne} P. Abreu et al., DELPHI coll.,
%                  Eur. Phys. J. {\bf C13} (2000) 591.
\bibitem{detec} P. Abreu et al., DELPHI coll., Nucl. Inst. and Meth.
                {\bf A378} (1996) 57;\\
                P. Aarnio et al., DELPHI coll., Nucl. Inst. and Meth.
                {\bf A303} (1991) 233.
\bibitem{bdk} F.A. Berends, P.H. Daverveldt and R. Kleiss,
              Comput. Phys. Commun. {\bf 40} (1986) 271.
\bibitem{wph} E. Accomando and A. Ballestrero,
              Comput. Phys. Commun. {\bf 99} (1997) 270;\\
              E. Accomando, A. Ballestrero and E. Maina, hep-ph/0204052,
              to appear in Comput. Phys. Commun.
\bibitem{kk2f} S. Jadach, B.F.L. Ward and Z. W\c{a}s,
               Comput. Phys. Commun. {\bf 130} (2000) 260.
\bibitem{bhw} S. Jadach, W. Placzek and B.F.L. Ward,
              Phys. Lett. {\bf B390} (1997) 298.
\bibitem{kora} S. Jadach, B.F.L. Ward and Z. W\c{a}s,
               Comput. Phys. Commun. {\bf 79} (1994) 503.
\bibitem{pyt} T. Sj\"ostrand,
              Comput. Phys. Commun. {\bf 39} (1986) 347.
%\bibitem{sgv} M. Berggren, R. Keranen, A. Sopczak,
%              Eur. Phys. J. direct {\bf C8} (2000) 1.
\bibitem{sgv} R. Keranen et al.,
              Eur. Phys. J. direct {\bf C7} (2000) 1.
              %{\tt http://berggren.home.cern.ch/berggren/sgv.html}
\bibitem{luclus}  T. Sj\"ostrand,
                  Comput. Phys. Commun. {\bf 82} (1994) 74.
\bibitem{durham} S. Catani et al.,
                 Phys. Lett. {\bf B269} (1991) 432.
\bibitem{lep1} P. Abreu et al., DELPHI coll.,
               Eur. Phys. J. {\bf C16} (2000) 371.
\bibitem{vo} V.F. Obraztsov,  Nucl. Inst. and Meth.
                {\bf A316} (1992) 388 and erratum {\bf A399} (1997) 500.
\bibitem{cousins} R.D. Cousins and V.L. Highland,  Nucl. Inst. and Meth.
                {\bf A320} (1992) 331.
\end{thebibliography}
\end{document}